\documentclass[twocolumn]{aastex631}
\usepackage{amsmath}
\usepackage{graphicx}
\usepackage{textcomp}
\usepackage{gensymb}
\usepackage{natbib}

\begin{document}
\def\teff{$T_{\rm eff}$}
\def\av{$A_{V}$}
\def\msun{$M_{\odot}$}

\graphicspath{{figures/}}

\received{}
\revised{}
\accepted{7 May 2024}
%% Command to document which AAS Journal the manuscript was submitted to.
%% Adds "Submitted to " the argument.
%\submitjournal{}

\shorttitle{Starspots and Binaries in Upper Sco}
\shortauthors{Sullivan \& Kraus}

\title{Starspots and Undetected Binary Stars Have Distinct Signatures in Young Stellar Associations}

\author[0000-0001-6873-8501]{Kendall Sullivan}
\affil{Department of Astronomy and Astrophysics, University of California Santa Cruz, Santa Cruz CA 95064, USA}

\author{Adam L. Kraus}
\affil{Department of Astronomy, University of Texas at Austin, Austin TX 78712, USA}

\correspondingauthor{Kendall Sullivan}
\email{ksulliv4@ucsc.edu}

\begin{abstract}
Young stars form in associations, meaning that young stellar associations provide an ideal environment to measure the age of a nominally coeval population. Isochrone fitting, which is the typical method for measuring the age of a coeval population, can be impacted by observational biases that obscure the physical properties of a population. One feature in isochrone fits of star-forming regions is an apparent mass-dependent age gradient, where lower-mass stars appear systematically younger than higher-mass stars. Starspots and stellar multiplicity are proposed mechanisms for producing the mass-dependent age gradient, but the relative importance of starspots versus multiplicity remains unclear. We performed a synthetic red-optical low-resolution spectroscopic survey of a simulated analog to a 10 Myr stellar association including mass-dependent multiplicity statistics and age-dependent starspot coverage fractions. We found that undetected starspots alone do not produce an apparent mass-dependent age gradient, but instead uniformly reduce the average measured age of the population. We also found that binaries continue to produce an apparent mass-dependent age gradient, and introduce more scatter in the age measurement than spots, but are easily removed from the population as long as there are good distance measurements to each target. We conclude that it is crucial to incorporate treatments of both starspots and undetected stellar multiplicity into isochrone fits of young stellar associations to attain reliable ages.
\end{abstract}

\section{Introduction}
Stars typically form in associations within giant molecular clouds \citep{Lada2003}. Determining the ages of these nominally coeval young stellar associations is important for a variety of scientific goals. For example, ages are used to calibrate evolutionary models \citep[e.g., MIST;][]{Paxton2011, Paxton2013, Paxton2015, Choi2016, Dotter2016}, determine the duration and detailed history of star formation \citep[e.g.,][]{Kerr2021}, reassemble star-forming events after the fact \citep[e.g.,][]{Zucker2022}, and measure the protoplanetary disk dissipation timescale, which sets the timescale for exoplanet formation \citep[e.g.,][]{Mamajek2009}. One of the most efficient techniques for measuring stellar ages is isochrone fitting, which compares sources on an HR diagram to evolutionary model tracks at various ages. Because young stars are still descending the Hayashi track \citep{Hayashi1961}, it is possible to place them on isochrones to determine their ages with high relative precision.

One key site of local star formation is the Scorpius-Centaurus (Sco-Cen) OB association, which is the closest collection of massive stars and thus also the one of the largest nearby associations of roughly coeval low-mass stars. Sco-Cen is comprised of three main subgroups: Upper Scorpius (or Upper Sco), Lower Centaurus-Crux, and Upper Centaurus-Lupus. Upper Sco is the youngest region, although there is substantial substructure with age spreads in each of the regions \citep[e.g.,][]{Kerr2021, Zerjal2023}. Sco-Cen hosts $\sim10^{4}$ low-mass stars \citep{Rizzuto2015}, and falls at a distance of $\sim 140$ pc \citep[e.g.,][]{Rizzuto2015, Zerjal2023}, making it an excellent location for studying young stellar populations, reconstructing the star formation history of stellar associations, calibrating evolutionary models, and measuring the circumstellar disk dissipation timescale.

However, one major barrier in using young stellar associations to study star and planet formation is difficulty in measuring their ages. In Upper Sco, early work \citep[e.g.,][]{Preibisch1999, Preibisch2001, Preibisch2002} suggested a 5 Myr age using primarily low-mass stars, while more recent work has supported a 10 Myr age using higher-mass stars \citep[e.g.,][]{Pecaut2012, Feiden2016, Pecaut2016, Sullivan2021}. Some work has observed an apparent mass-dependent age gradient in Upper Sco, where low-mass stars almost always appear younger than high-mass stars even in a uniformly observed and characterized population \citep{Pecaut2012, Rizzuto2015, Pecaut2016}, suggesting that disagreement between different analyses could be caused by a systematic feature of the population. If the observed mass-dependent age gradient is physical, it could indicate hierarchical star formation (with massive stars forming after low-mass stars; \citealt{Elmegreen1977, Preibisch1999, Krumholz2019}). However, most work has focused on potential observational biases or astrophysical features of young stellar populations that could produce an apparent mass-dependent age gradient in an approximately coeval population. Some of the possible observational or astrophysical sources of this systematic feature in associations are starspots, undetected binary stars, uncertain distances, uncertain extinctions, and multimodal star formation. Each of these phenomena has an impact on the inferred HR diagram, and can alter age measurements. 

Much of the focus on systematic effects that can alter inferred isochronal ages has fallen on magnetic activity, especially starspots \citep{Somers2015, Feiden2016, Gully-Santiago2017, Somers2020, Cao2022a, PerezPaolino2023, PerezPaolino2024}. Evolutionary models incorporating starspots can more accurately reproduce observed radii of young eclipsing binaries \citep[e.g.,][]{Tofflemire2023}, and young stars are known to be heavily spotted \citep[e.g.,][]{Gully-Santiago2017, Morris2020, Cao2022b, PerezPaolino2023}. The presence of starspots introduces a secondary temperature component into the stellar spectrum, which can cause the inferred stellar temperature to be cooler than the photospheric temperature \citep[e.g.][]{Gully-Santiago2017, PerezPaolino2024}. Because starspot coverage fraction is dependent on age and magnetic activity level \citep[e.g.,][]{Morris2020, Somers2020}, undetected starspots could produce an apparent mass-dependent age gradient that may be similar to the gradient observed in Upper Sco. Magnetic activity can also impact isochrone fits by altering the interior structure of the star \citep{Feiden2016}, and the \citet{Feiden2016} analysis of Upper Sco that incorporated magnetic fields (but not starspots) found a consistent 10 Myr age for Upper Sco across all stellar masses. The magnitude of the effect of starspots on observations is not established, and it is unclear if starspots alone sufficiently explain the observed mass-dependent age gradient in Upper Sco.

An alternative explanation for the observed mass-dependent age gradient of Upper Sco is undetected binary stars. Binaries can produce a similar observational effect to starspots, with the presence of a secondary star introducing a cooler component into the spectrum. \citet{Sullivan2021} simulated an Upper Sco analog without starspots but including undetected binary stars, and found that undetected binaries in a single-age 10 Myr population caused low-mass stars to appear several Myr younger than high-mass stars. They also found that undetected binaries introduce significant scatter into the HR diagram that increases toward lower masses, which is expected given mass-dependent binary statistics.

One key distinguishing feature between starspots and undetected binaries in an HR diagram is that starspots reduce the luminosity of a system while decreasing the apparent temperature, so systems with spots will appear fainter and cooler, while binaries increase the luminosity of a system while decreasing the apparent temperature, so systems with binaries will appear brighter and cooler. Stellar evolution leads to stars becoming less luminous at the same \teff\ as they age, meaning that a brighter, cooler star will appear younger than a fainter, cooler star, because the spotted star will more closely follow its original isochrone (see Figure \ref{fig:single_isochrone} for an illustration of this effect). However, distinguishing between these scenarios requires an accurate distance to measure a precise luminosity, which was difficult for analyses before the \textit{Gaia} mission \citep{Gaia2016, Gaia2018, Gaia2021, Gaia2022}, or for extincted, optically faint populations. Another complicating factor is the possible multimodal star-formation history of Upper Sco \citep[e.g.,][]{Zerjal2023}, but a multi-age population should only introduce additional scatter on top of that caused by undetected binaries and starspots, rather than introducing additional features like the mass-dependent age gradient.

To expand on the work of \citet{Sullivan2021} and explore the relative impact of starspots versus undetected binary stars on isochronal ages, we simulated a spectroscopic survey of an analog of the Upper Scorpius star-forming region including both undetected binary stars and starspots using the SPOTS evolutionary tracks \citep{Somers2020}. In this paper we present the results from simulated age and mass measurements of spotted single stars at an age of 10 Myr, as well as results from a similar simulation that incorporated both spots and unresolved binaries.

\section{Simulating a Spectroscopic Survey of Upper Scorpius}
Our original simulated survey of an analog of Upper Scorpius including unresolved binaries is described in \citet{Sullivan2021}. In the following subsections we provide a brief summary of our original methodology and detail the changes made to the simulation for this work. 

\subsection{Population Statistics}
We assigned properties to our systems at the population level using an initial mass function (IMF), observed binary star properties, and estimated starspot properties. For the desired number of systems, we drew stellar masses using a Chabrier IMF \citep{Chabrier2003, Chabrier2005}, which is defined as a Salpeter power law \citep{Salpeter1955} at the high-mass end and a lognormal on the low-mass end:

\[
	\xi(\log(m)) \propto
	\begin{cases}
		\exp(\frac{(\log(m) - \log(m_{c})^{2}}{2  \sigma^{2}}) & \text{if } m \leq M_{cutoff}\\
		m^{-\alpha} & \text{if } m >  M_{cutoff}\\
	\end{cases}
\]

where we used the parameters from \citet{Chabrier2005} and thus defined $m_{c} = 0.2$\msun, $\sigma = 0.55$\msun, $M_{\text{cutoff}} = 1$\msun, and $\alpha = 1.3$. We artificially reduced the number of low-mass stars at an arbitrary second cutoff mass to produce more high-mass stars with a smaller population size, but performed that reduction after creating the initial IMF. A typical second cutoff mass was 0.8\msun, where stars below that mass were suppressed by a factor of 25. This reduction did not change our analysis or conclusions, and only served to increase our computational efficiency.

\begin{figure*}
\plotone{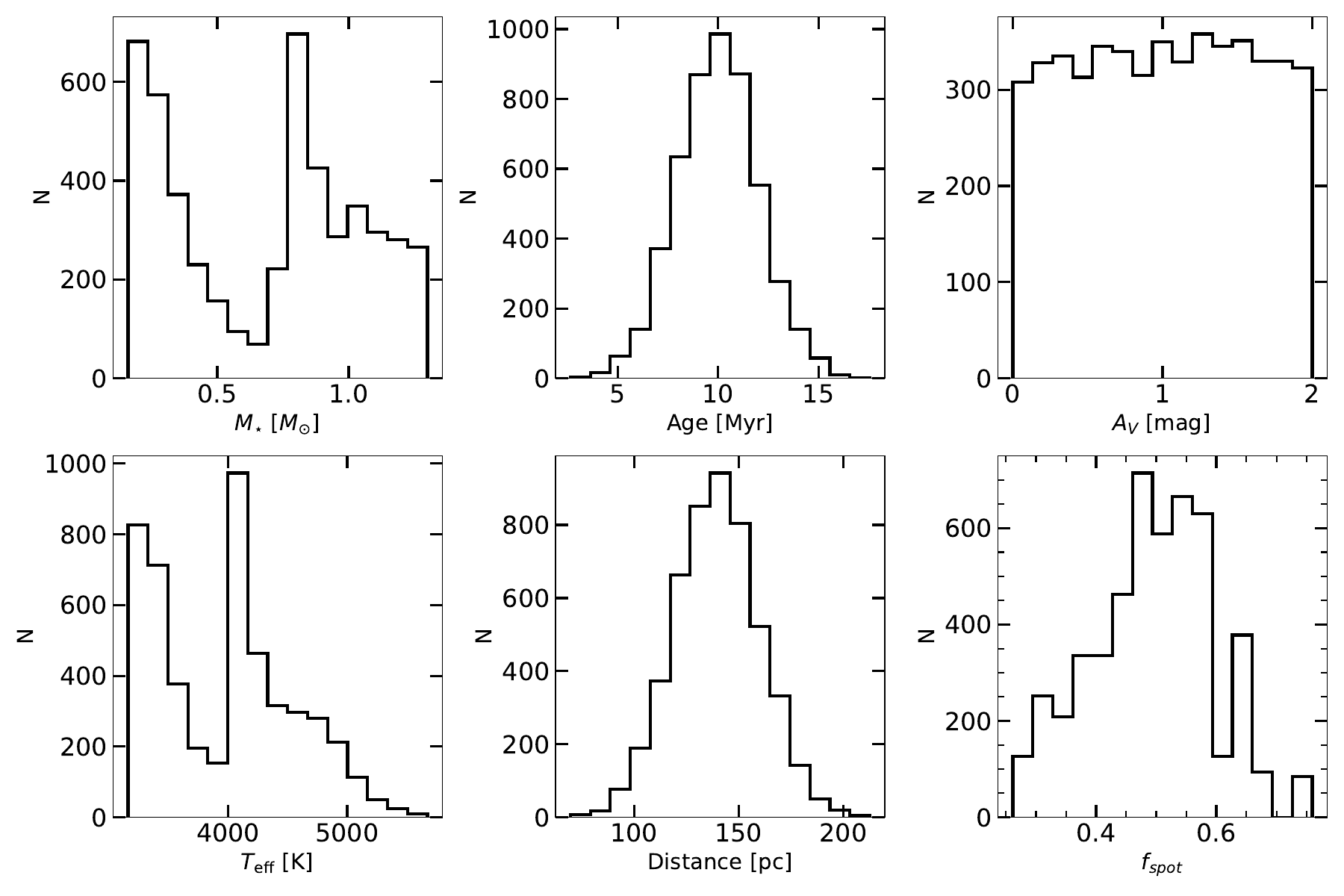}
\caption{Histograms for the simulated input population of single stars. Top row: stellar mass, age, and extinction. Bottom row: stellar \teff, distance, and spot coverage fraction. The apparent bimodality in mass, \teff, and spot coverage fraction is a result of the suppression of low-mass stars below 0.8\msun\ to improve computational efficiency.}
\label{fig:primary_stats}
\end{figure*}

After drawing a primary star mass, we assigned an age to each system by randomly drawing a value from a normal distribution with $\mu = 10$Myr and $\sigma = 2$Myr. These ages matched the age distribution in \citet{Sullivan2021} and were consistent with the age derived using F stars in \citet{Pecaut2012}. We also assigned a distance for each system, which we drew from a normal distribution with $\mu = 140$pc and $\sigma = 20$pc, based on results from \citet{Rizzuto2015}. With the assigned distance, mass, and age of each primary star, we added starspots and stellar multiplicity to create a complete system. Figure \ref{fig:primary_stats} shows various distributions from the simulated primary star population.

\begin{figure*}
\plotone{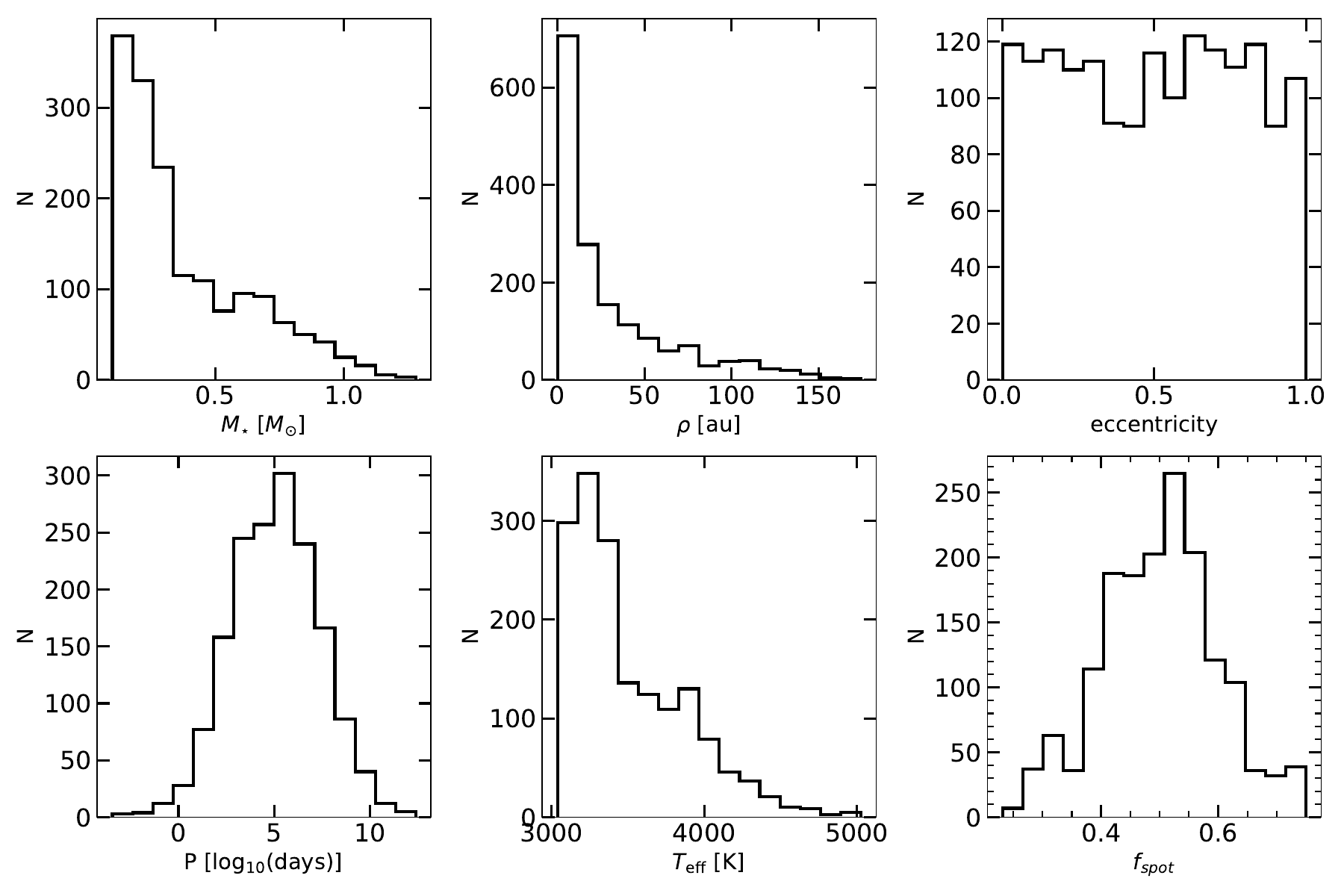}
\caption{Histograms for the simulated input population of secondary stars. Top row: secondary star mass, separation, and eccentricity. Bottom row: period, secondary star \teff, and starspot coverage fraction.}
\label{fig:secondary_stats}
\end{figure*}

To add stellar multiplicity, we assigned a system a secondary star based on its mass using a piecewise mass-dependent multiplicity fraction function of $f = \{0.25, 0.35, 0.45, 0.7\}$ with transitions between values at transition masses $m = \{0.15, 0.4, 1.2\}$\msun. These fractions were based on observations from the literature \citep[e.g.,][]{Kraus2008, Raghavan2010, Duchene2013, DeRosa2014, Ward-Duong2015, Winters2019, Tokovinin2020}. If a system was a multiple, we assigned it a separation and mass ratio that were again based in observations and mass-dependent, using functions derived in \citet{Sullivan2021}. We assigned a resolution limit of 1'', and discarded any secondary star that fell outside that separation limit, since our goal was to investigate unresolved and undetected binary stars. An arcsecond is approximately the spatial resolution of the Gaia mission \citep{Gaia2016} and is also a typical value for good seeing at most ground-based observatories, meaning that many binaries closer than 1'' would not be detected without dedicated surveys to detect multiplicity. The properties of the generated secondary star population are shown in Figure \ref{fig:secondary_stats}.

To add starspots, we imposed a normally distributed spot coverage fraction with a mean value of 50\% and a standard deviation of 10\%. For drawn values below 0\% or above 80\% we resampled the distribution to avoid nonphysical values or starspot covering fractions that were outside the range of the SPOTS models \citep{Somers2020}. Only a small fraction of systems ($< 1$\%) needed to be redrawn because they fell outside the limits.

\subsection{Data Generation}
Using the statistics described above, we produced systems with the following known properties: mass, age, distance, and starspot coverage fraction, as well as mass ratio, separation, and a second starspot coverage fraction if the system was a binary. To produce simulated data using model spectra, we had to convert these physical properties to observables using evolutionary models. We used the SPOTS models \citep{Somers2020}, which incorporate starspots and provide both starspot and photospheric temperatures. The SPOTS models have a fixed range of possible spot coverage fractions, ranging from 0\% to 85\%, and a limited mass range of 0.1-1.3\msun. Therefore, for this analysis, we restricted the simulated population to primary star masses $0.16 \leq M_{p} \leq 1.3$ \msun, and we redrew any secondary stars with masses that fell below 0.1 \msun. Since the mass ratio distribution of low-mass binaries becomes increasingly biased toward unity \citep[e.g.,][]{Winters2019}, the number of secondaries that needed to be redrawn was small.

\begin{figure}
    \plotone{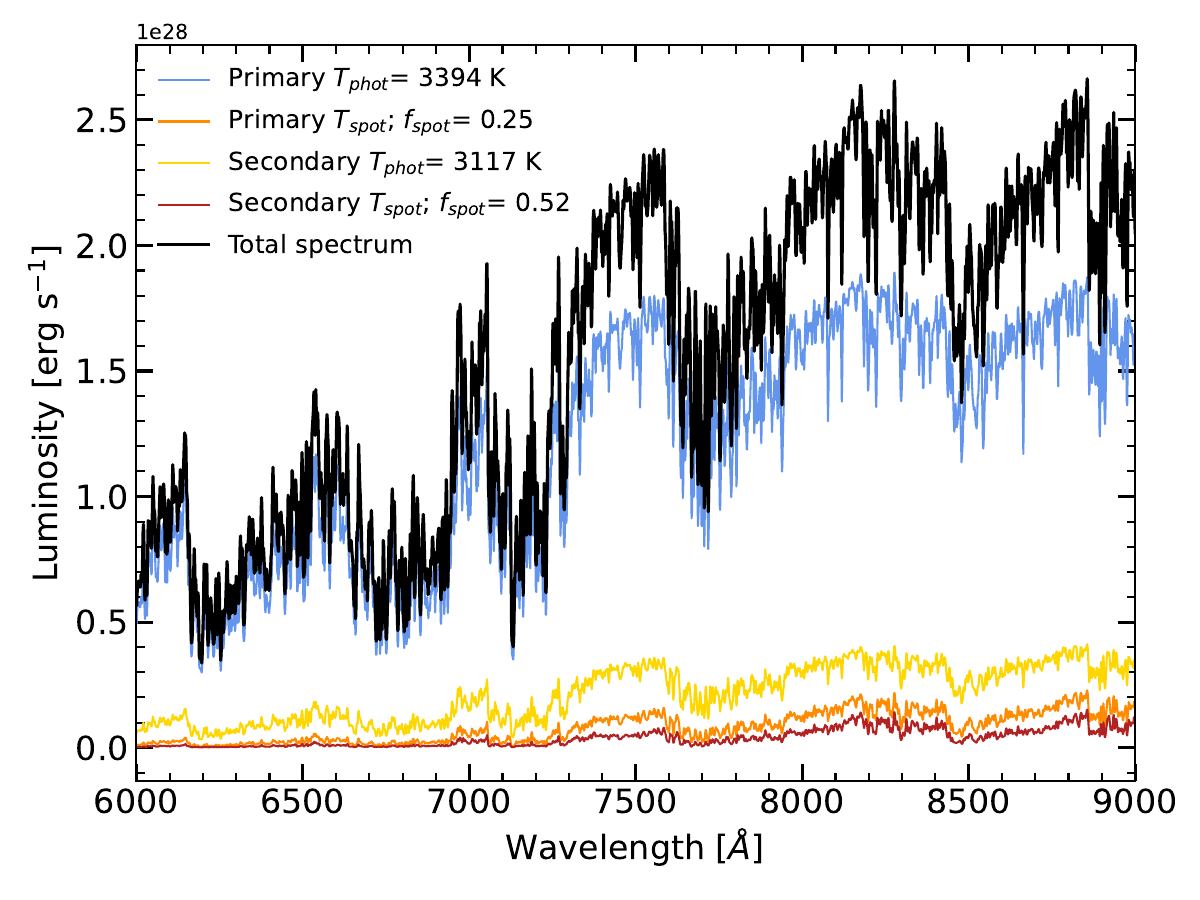}
    \caption{Composite and component spectra for a typical binary system. The composite spectrum (black), primary photosphere (blue), primary spot (orange), secondary photosphere (yellow) and secondary spot (red) spectra are all shown. The primary flux dominates at the short wavelengths, but the cooler components progressively contribute more flux toward longer wavelengths.}
    \label{fig:components}
\end{figure}

Using the SPOTS models, we linearly interpolated the models in spot coverage fraction, mass, and age, and calculated a photospheric temperature, spot temperature, luminosity, surface gravity, and VRIJHK magnitudes for each component of the system. We used the default SPOTS spot temperature contrast value of $T_{spot}/T_{phot} = 0.8$. Using the luminosity and temperature, we inferred a stellar radius ($R_{\star} = \sqrt{\frac{L_{\star}}{4 \pi \sigma T_{\star}^{4}}}$) for the primary and secondary. We created a spectrum for each component of each photosphere (i.e., model spectra at T$_{phot}$ and T$_{spot}$ for singles) by bilinearly interpolating each spectrum pixel-by-pixel in \teff\ and $\log(g)$ using the \citet{Husser2013} model spectra calculated with the PHOENIX code. We chose these model spectra for their wide range of \teff\ and surface gravity. We normalized each component to the appropriate surface area (e.g., $A_{T_{1, phot}} = R_{1}^{2} * (1 - f_{spot, 1})$, where $f_{spot, 1}$ is the spot coverage fraction), then summed the two (for singles) or four (for binaries) spectra (e.g., Figure \ref{fig:components}) to produce a single composite spectrum for each system. To match previous moderate-resolution studies of Upper Sco, we created the spectra at a resolution of R$\sim$2100 by convolving with a Gaussian of the appropriate width, and covered the red-optical wavelength range from 6000-9000 \AA\ \citep[e.g.][]{Preibisch2002, Pecaut2012, Rizzuto2015, Pecaut2016}.

We also analyzed SEDs for each system created by synthetic broadband photometry. The SPOTS models already incorporate the impact of spots into the model photometry, so we simply added the distance modulus and combined magnitudes for the two components of the binaries by summing the fluxes. We assumed a distance measurement with a typical parallax error of 0.05 mas (consistent with \textit{Gaia} parallax measurements for relatively bright sources; \citealt{Gaia2021}), so we inverted each assigned system distance, perturbed it with a Gaussian with $\sigma = 0.05$ mas, then inverted the perturbed parallax to achieve a perturbed distance ``measurement''. Finally, we extincted each component with a randomly assigned extinction ranging from $0 \leq A_{V} \leq 2$ drawn from a uniform distribution, and converted between wavelengths using the extinction law from \citet{Cardelli1989}. At the conclusion of the data generation process we had a single spectrum and SED in VRIJHK for each system, incorporating starspots and undetected binaries with separations $\rho < 1$''. 

\subsection{System Analysis}
To analyze each composite spectrum, we used a Markov Chain Monte Carlo (MCMC) method implemented as a modified Gibbs sampler to retrieve a temperature for each system, then used the VRIJHK SED, best-fit \teff\ and \av\, and assigned distance to measure the luminosity. Using the best-fit luminosity and temperature, we inferred a mass and age. 

To fit each system, we initialized a walker with a guess \teff, surface gravity, and \av\ defined by the initial (primary star) values. To avoid beginning in an exactly correct location for the single stars, we perturbed the initial \teff\ and \av\ using a normal distribution with $\sigma_{T_{\rm eff}} = 100$ K and $\sigma_{A_{V}} = 0.1$ mag. We also fit for a normalization factor, which we initially perturbed by $\sigma_{\rm norm} = 1$\%, to account for the extra flux introduced by the secondary star. The normalization factor is required because we flux-normalized the ``observed'' spectra, rather than flux-calibrating the model spectra. We found that at the relatively low spectral resolution of our data our fit was not very sensitive to surface gravity changes, so we did not perturb the initial $\log{g}$ guess, and we did not fit for a new surface gravity value. Our fit results were not sensitive to the choice of initialization values, as discussed in \citet{Sullivan2021}. To perform the modified Gibbs sampling, we took the perturbed guess values, produced a single-star spectrum using those values, and compared the model to the data using a $\chi^{2}$ value. If the $\chi^{2}$ was lower than the previous best value, we always accepted the guess, and if the $\chi^{2}$ was higher we always rejected the guess. In contrast, a traditional Gibbs sampler occasionally accepts worse guesses than the current recorded best fit.

We fit in two stages, each of which had a coarsely sampled optimization fit followed by a fit that was more finely sampled to achieve the best-fit parameter values. In the initial coarse fit we drew random guesses from relatively broad normal distributions ($\sigma_{T_{\rm eff}} = 100$ K, $\sigma_{A_{V}} = 0.05$ mag, $\sigma_{\rm norm} = 1$\%) until the fitter ran for half the requested number of steps without finding a better fit. We then switched to finer sampling, using narrow normal distributions ($\sigma_{T_{\rm eff}} = 5$ K, $\sigma_{A_{V}} = 0.005$ mag, $\sigma_{\rm norm} = 0.05$\%) until the fitter ran for the other half of the requested number of steps without finding a better fit. We implemented an initial ``burn-in'' fit of 40 steps (i.e., a fit that reached 40 steps with one guess without achieving a better fit), then a full fit that ran for 200 steps. The typical reduced-$\chi^{2}$ values after the second fitting stage were $\chi^{2}<<1$ for single stars without spots, which was expected given that we did not introduce additional artificial random noise, meaning that in the ideal case (i.e., an infinite run time) the reduced-$\chi^{2}$ for a single star trend toward zero. We tested adding artificial noise to more closely match real observations, but found that it did not change our results and substantially slowed down our fits, so we chose to run our analysis in the optimal zero-noise scenario. 

Our fitting method retrieved a best-fit composite \teff, \av, and normalization value for each system. Validation of this technique on single stars and tests of the extinction fitting are described in \citet{Sullivan2021}. In general, for single stars, we were able to retrieve temperatures within a few K, and extinctions within 0.1 mag, although the fits became worse at higher temperatures (above $\sim 4500$ K) because of a reduction in the number of spectral features present in our wavelength range at low spectral resolution. However, because our young stars were almost all relatively low-mass, almost all of our stars were expected to have precise fits in the spot-free single star scenario.

We also fit the VRIJHK SED to measure a composite luminosity for each system using the MESA Isochrones and Stellar Tracks (MIST) stellar evolutionary models \citep{Paxton2011, Paxton2013, Paxton2015, Choi2016, Dotter2016}. Using the best-fit temperature, we interpolated the luminosity at that temperature as a function of each point in the SED, then took the median of the interpolated luminosities as the best-fit value. Using the luminosity and temperature, we also measured an age and a mass for each composite system. Because the MIST evolutionary models do not go below 3000 K, we removed any systems that had a best-fit \teff\ cooler than 3000 K from the final results. This step typically eliminated fewer than 100 systems from the total sample of $\sim 10^{4}$ systems, a decrease of 1\% of our total sample size at most.

\section{Simulation Results}\label{sec:results}
We simulated two populations, with one run composed of only single stars with starspots and the other run composed of single and undetected binary stars, all with starspots. Both runs can be compared against \citet{Sullivan2021}, with simulations for single stars and binaries without spots. The single-star-only simulation provided a benchmark for the expected outcome if spots were the dominant factor in producing scatter or apparent mass-dependent age gradients in a single-age population, while the simulation including single and binary stars allowed a comparison between the impact of undetected multiplicity and the impact of starspots. Each run consisted of 5000 simulated systems with primary star masses ranging from 0.16 \msun\ to 1.3\msun. 

\subsection{Single Star Simulation}\label{sec:results_single}

\begin{figure*}
    \plottwo{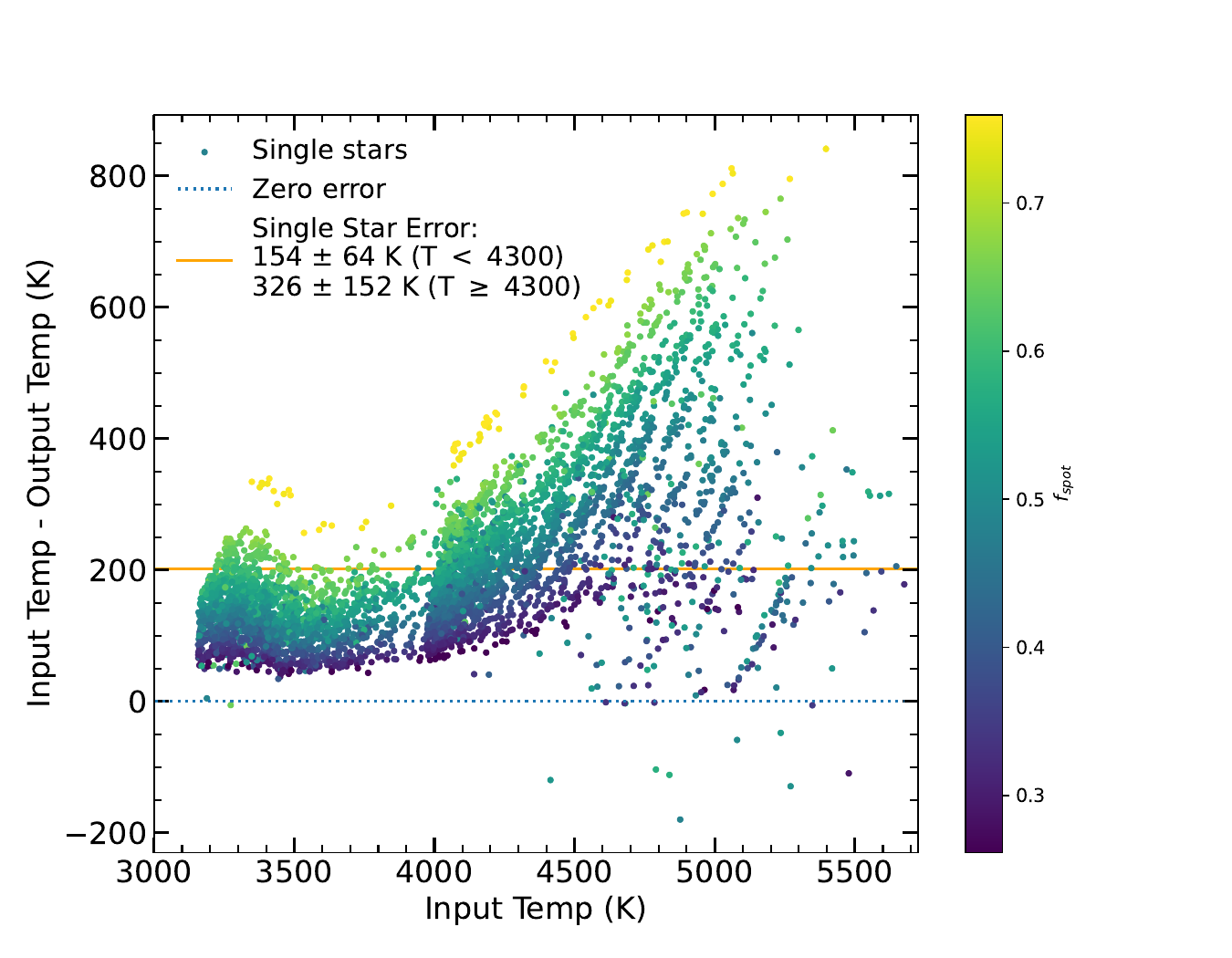}{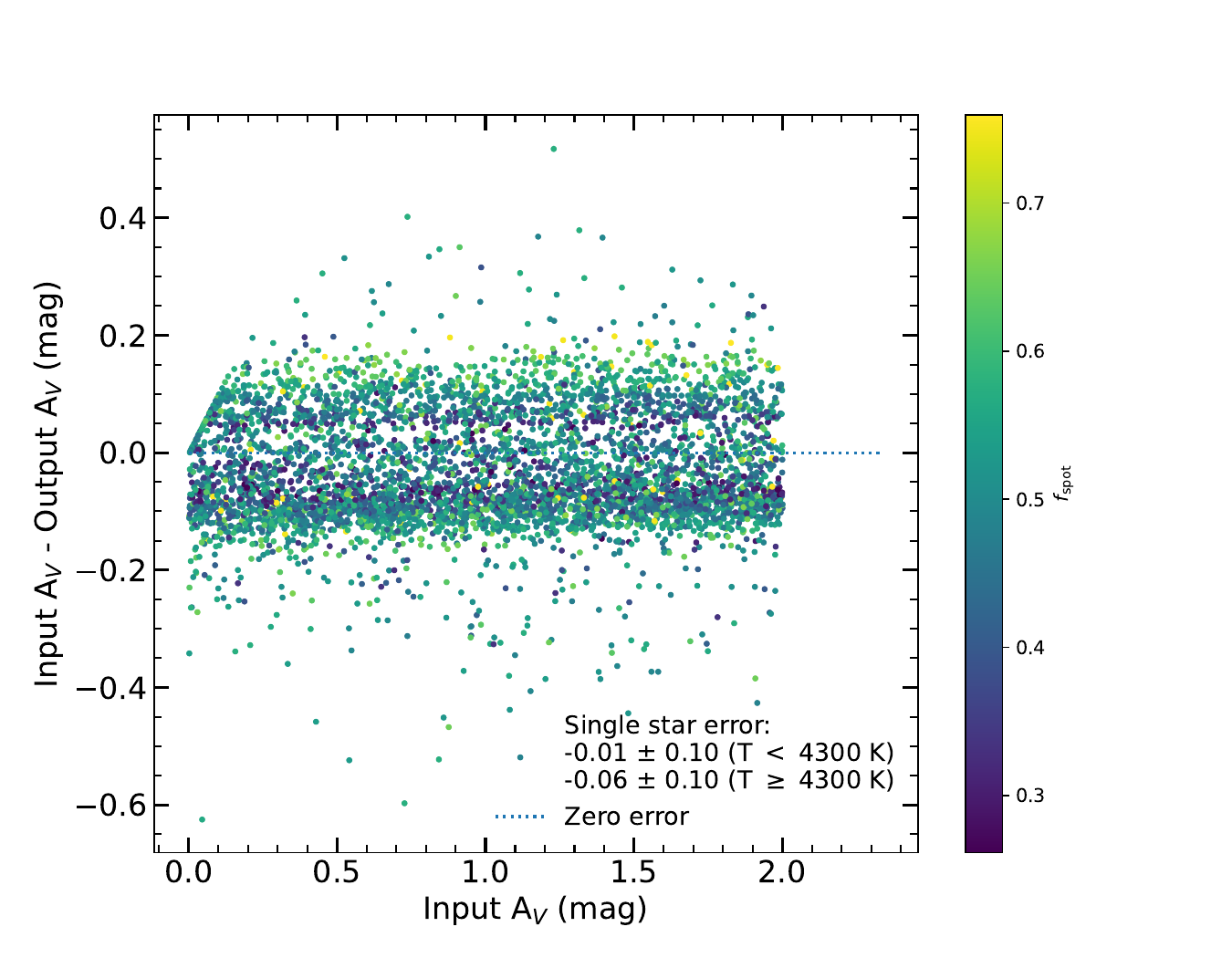}
    \caption{Fundamental retrieved parameters for single stars with spots. Left: \teff\ residual plotted against the input (true) primary star \teff. Cooler stars (M and late K) have larger systematic \teff-dependent offsets because of spots than hotter stars (early K and G). The average \teff\ residual is 154 (rms = 64) K (\teff $<$ 4300 K) and 326 (rms = 152) K (\teff\ $>$ 4300 K). The increased scatter in hotter stars results from a reduction in the number of \teff-sensitive spectral features at hotter temperatures with our spectral resolution and wavelength range. The measured \teff\ for stars with spots are typically systematically cooler than the true photospheric \teff. Right: \av\ residual versus input \av\ for all single stars with spots. The presence of spots does not bias the \av\ retrieval, with residuals of -0.01 (rms = 0.1) mag (\teff $<$ 4300 K) and -0.06 (rms = 0.1) mag (\teff $>$ 4300 K). The \av\ residual is correlated with $f_{spot}$ because of the reddening introduced by the presence of uncorrected spots. The gap in the residual around 0.03 is caused by the artificial reduction of low-mass stars, and is not reflective of any bias in the \av\ recovery.}
    \label{fig:teff_av_singles}
\end{figure*}

To demonstrate the impact of starspots  without the presence of binary stars, we simulated a population of single stars with spots. Figure \ref{fig:teff_av_singles} shows the \teff\ and \av\ residuals for single stars with spots after fitting the single-only, spotted stars using a single-\teff\ spectrum. The left panel includes the \teff\ residual plotted against the input photospheric temperature. Nearly all of the systems at all temperatures have a derived \teff\ that is cooler than the photospheric \teff, demonstrating the well-established impact of starspots on inferred stellar \teff\ \citep[e.g.,][]{Gully-Santiago2017, Cao2022a}. At lower \teff\ (\teff $<$ 4300 K), there is a systematic residual; the average residual is 154 (64) K. At slightly higher \teff\ (\teff $>$ 4300 K), there is an even larger systematic effect of spots, with an average residual of 326 (rms = 152) K. The scatter at low \teff\ is driven by the \teff\ bias produced by large starspot fractional coverage, while the scatter at high \teff\ is a result of spots, in addition to fewer \teff-sensitive spectral features being present at higher \teff\ at our chosen spectral resolution and wavelength range, as discussed in \citet{Sullivan2021}. 

In contrast to the \teff\ measurement, there is no systematic residual with temperature in \av, with residuals of -0.01 (rms = 0.1) mag for \teff $<$ 4300 K and -0.06 (rms = 0.1) mag for \teff $>$4300 K. This excellent \av\ recovery remains consistent with the results from \citet{Sullivan2021}, which found that \av\ was recovered accurately for both single stars and undetected binary stars in systems without spots. The \av\ residual is correlated with the spot coverage fraction, with a higher spot coverage fraction being correlated with an underestimated observed \av\ value. The reason for this underestimation is unclear, but is likely related to the presence of starspots, given that the same feature does not occur in the spot-free single star simulations in \citet{Sullivan2021}. The apparent gaps in the residual are caused by small gaps in the mass function produced by the artificial reduction in the number of low-mass stars, and are not a physical feature of the \av\ residual structure.

\begin{figure}
    \plotone{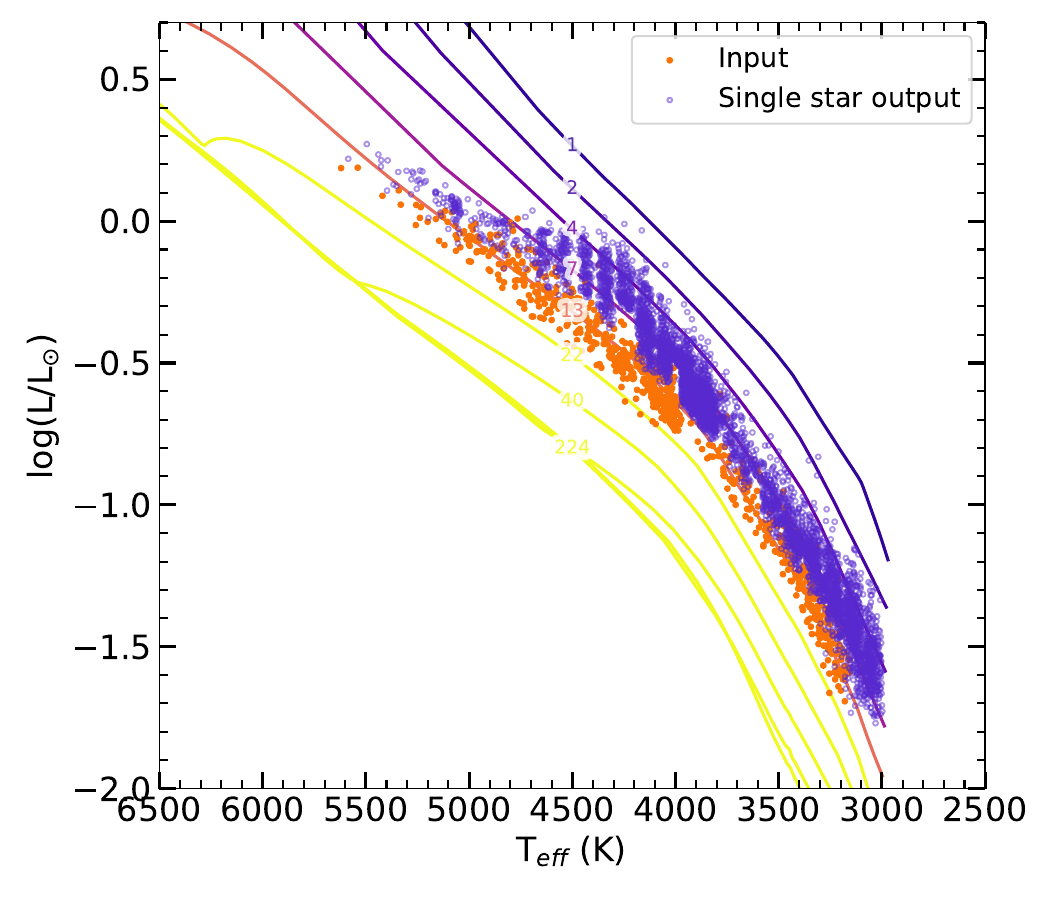}
    \caption{An HR diagram showing our input (orange) and measured (purple) \teff\ and luminosity for single stars with spots. The underlying curves are MIST isochrones. The input population's photospheric \teff\ and luminosities lie uniformly along the 10/13 Myr isochrone, while the output population is uniformly raised above the input isochrone because of the presence of starspots.}
    \label{fig:single_isochrone}
\end{figure}

To demonstrate the observational impact of starspots on derived ages, Figure \ref{fig:single_isochrone} shows an HR diagram of our input and output simulated populations with MIST isochrones underlaid. The input population (in orange) falls along the correct isochrone with no apparent gradient, while the output population appears younger and shows more scatter than the single population. 

\subsection{Binary Star Simulation}
To explore the relative impact of undetected binaries and undetected starspots, we simulated a population that was identical to the single-star-only simulation except for the inclusion of a population of undetected binary stars with separations $\rho < 1''$.

\begin{figure*}
    \plottwo{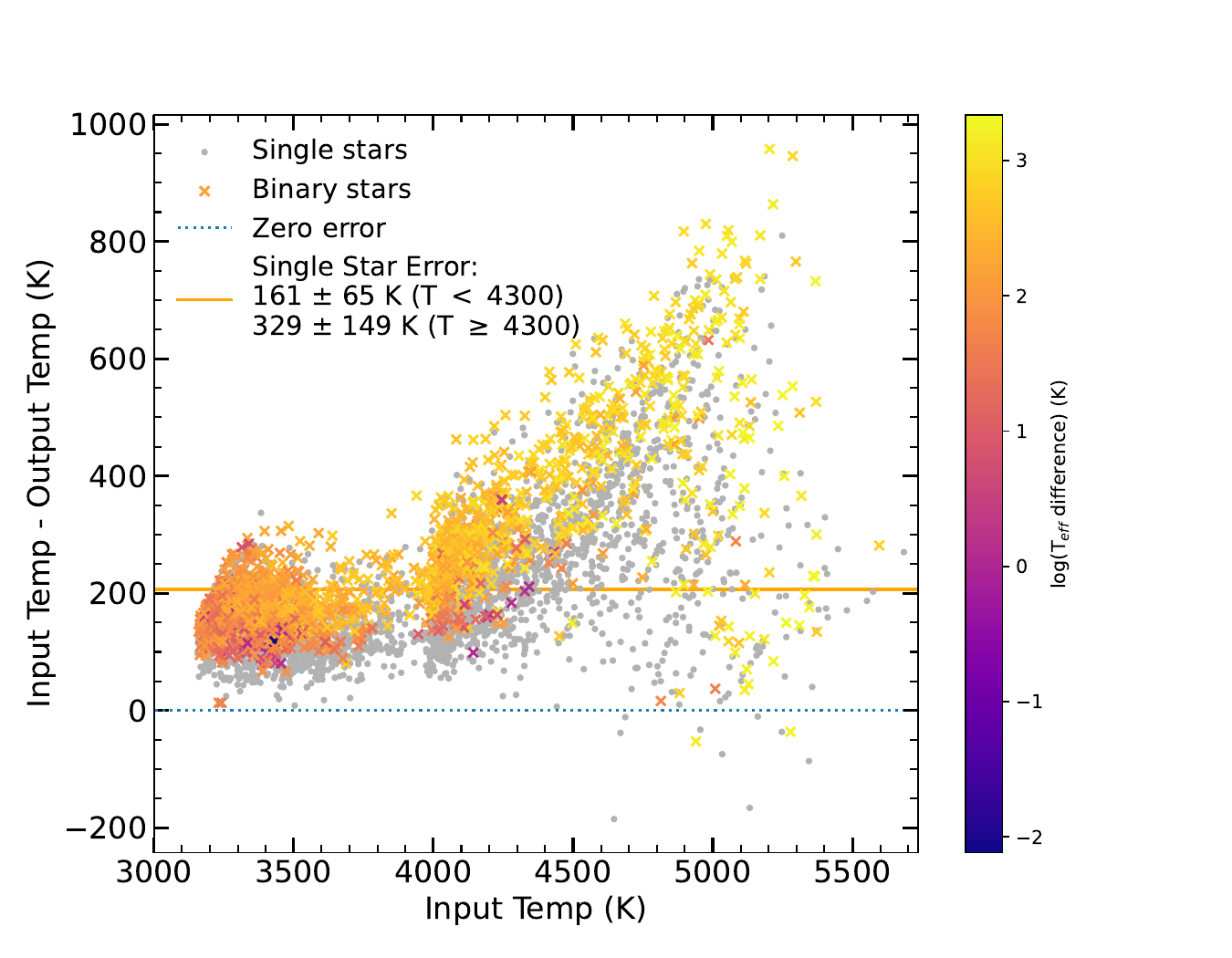}{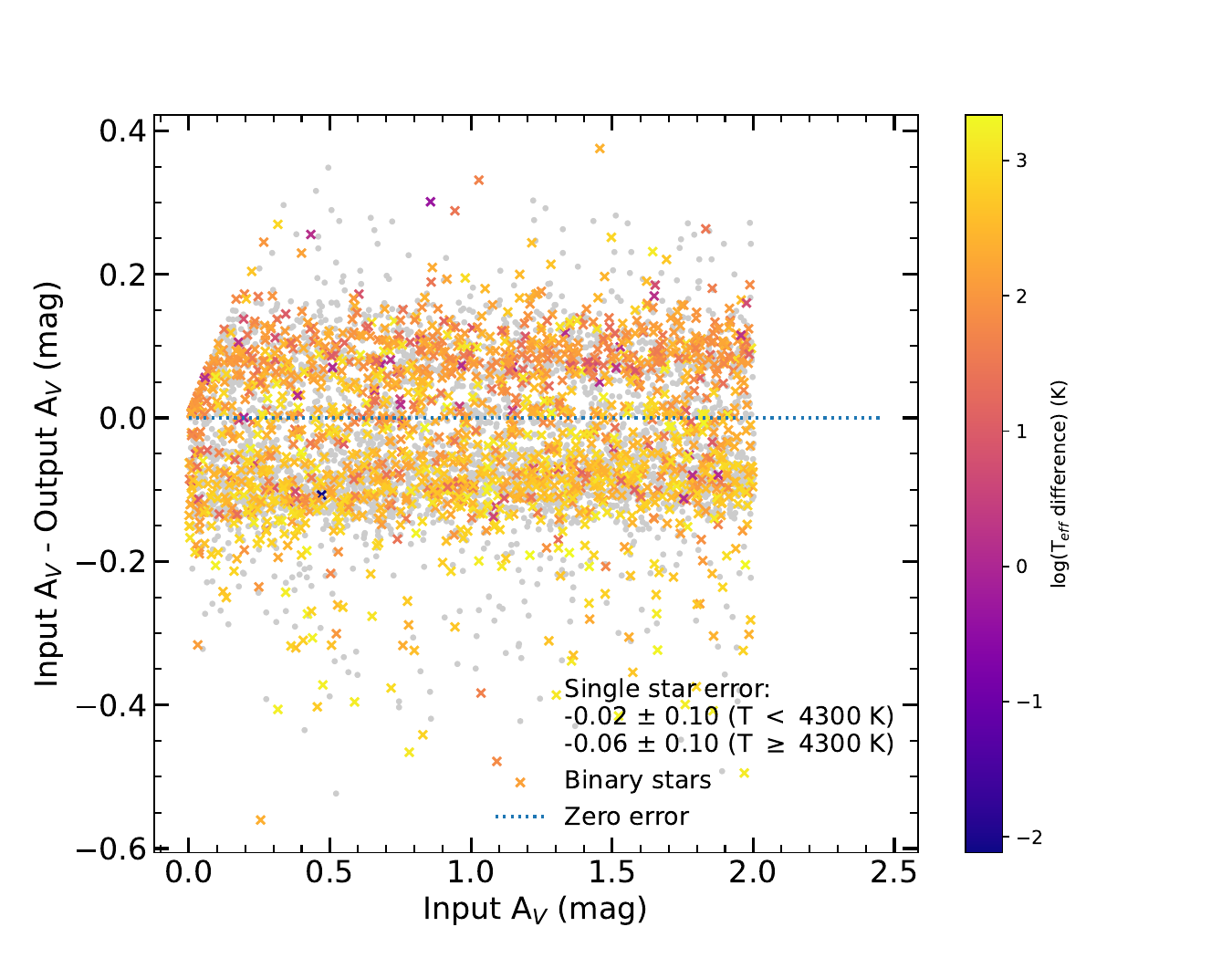}
    \caption{The same as Figure \ref{fig:teff_av_singles}, but for the simulation including unresolved binary stars (cross symbols, color-coded by \teff\ difference between primary and secondary star). Left: The single-star \teff\ residual is identical to the single-star-only run, as expected, while the systems hosting unresolved binary stars typically have \teff\ measurements that are biased toward even cooler temperatures, with typical residuals of $\sim$150 K for the cooler stars. Right: The \av\ measurement is not biased by the presence of binaries, demonstrated by the residuals that match those of the single-star-only case. However, for binaries, the residual is dependent on the \teff\ contrast between the two stellar components, likely because of the extra apparent reddening of the SED caused by the secondary star.}
    \label{fig:teff_av_binaries}
\end{figure*}

Figure \ref{fig:teff_av_binaries} shows the \teff\ and \av\ residuals plotted against the input \teff\ and \av\ values. The \teff\ residuals for single stars match those in the single-star-only simulation, which is expected because we did not change any of the population parameters for single stars. The binary stars systematically have larger \teff\ residuals than the single stars, with moderate-contrast binaries having the largest \teff\ residuals. This is because moderate-\teff-contrast systems have secondaries with sufficient additional flux and spectral morphology changes to bias the \teff\ measurement. Similar to the single stars, the binary bias decreases toward higher masses as the spot coverage fraction decreases, but then increases again as the scatter grows at earlier spectral types. Regardless of the input primary star photospheric \teff, the binaries consistently have larger \teff\ residuals than the single stars. 

In contrast to the \teff\ measurements, the \av\ residuals shown in Figure \ref{fig:teff_av_binaries} are uniformly distributed, although they do show a slight bias that is dependent on the \teff\ contrast between the components of the binary, unlike the results from \citet{Sullivan2021}. The impact on the \av\ measurement is not large (a maximum of $\sim$0.2 mag deviation from the true \av\ value), so we do not expect it to significantly impact our analysis or conclusions.

\begin{figure}
    \plotone{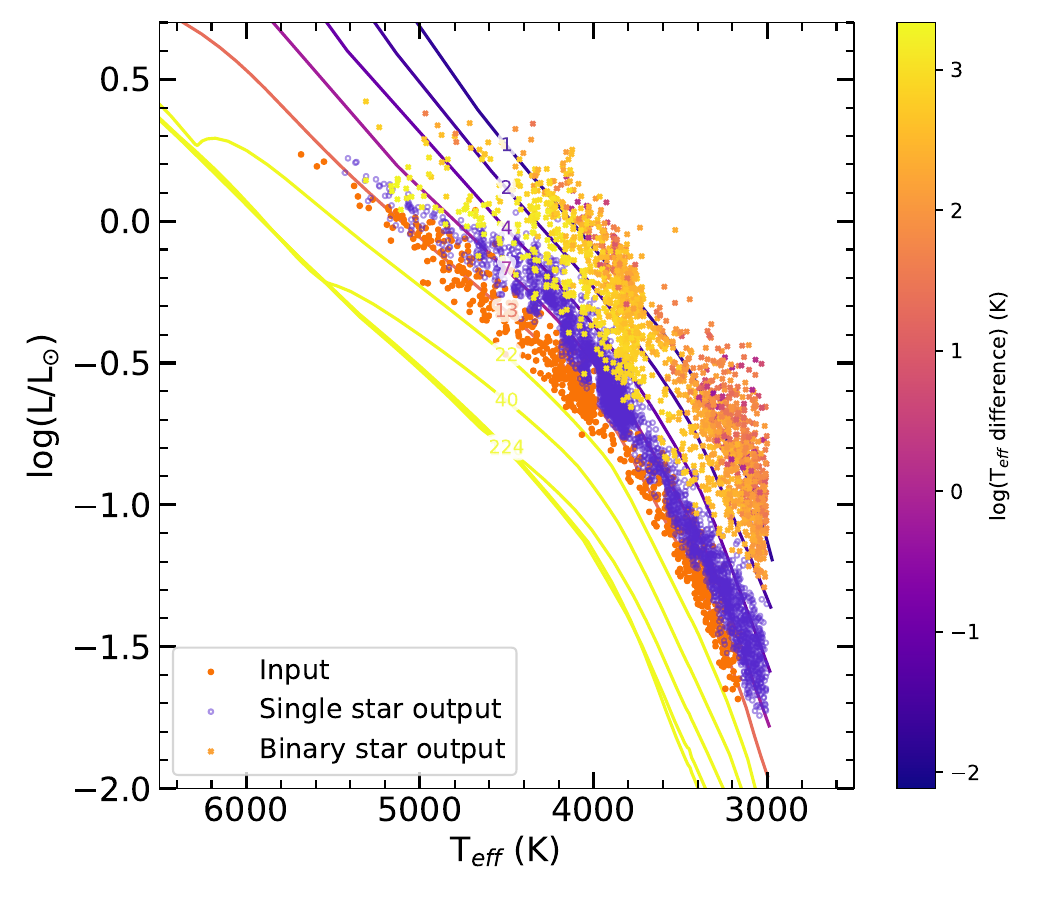}
    \caption{The same as Figure \ref{fig:single_isochrone}, but for the simulation including binary stars (shown as cross markers, with color coding indicating the \teff\ contrast between the primary and secondary star). The binary stars systematically fall at younger ages and higher luminosities than either the single star input or output populations, even with spots included. The binaries have a large amount of scatter and another apparent mass-dependent age gradient, both produced by the mass-dependent population statistics of binary stars.}
    \label{fig:binary_isochrone}
\end{figure}

To compare to a typical HR diagram of a star-forming region, we plotted the simulated population on an HR diagram along with MIST isochrones, shown in Figure \ref{fig:binary_isochrone}. The binary stars systematically fall on younger isochrones and with larger luminosities than the single star input or output populations. The binaries have a large amount of scatter and show a mass-dependent age gradient matching that seen in \citet{Sullivan2021}, where systems with hotter primary stars appear older than systems with cooler primaries. With the improved distance precision from \textit{Gaia} a binary sequence is apparent, with most binaries falling above the sequence of single stars. This is the result of spots causing single stars to appear fainter as well as cooler, while the binaries appear brighter and cooler, creating a more distinct separation than for systems without spots.

\section{Comparing Age Measurements Between Simulated Populations} 
\begin{figure*}
    \centering
      \includegraphics[width=0.49\textwidth]{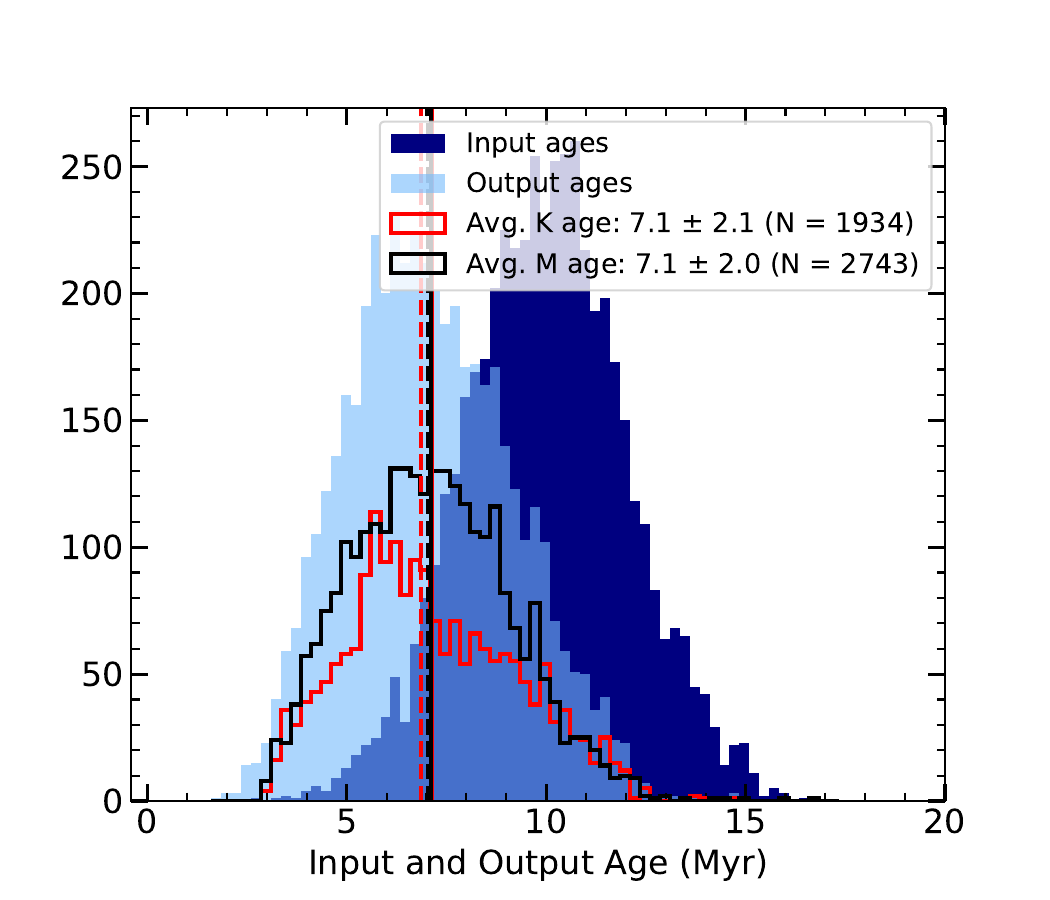}
    \\
      \includegraphics[width=0.49\textwidth]{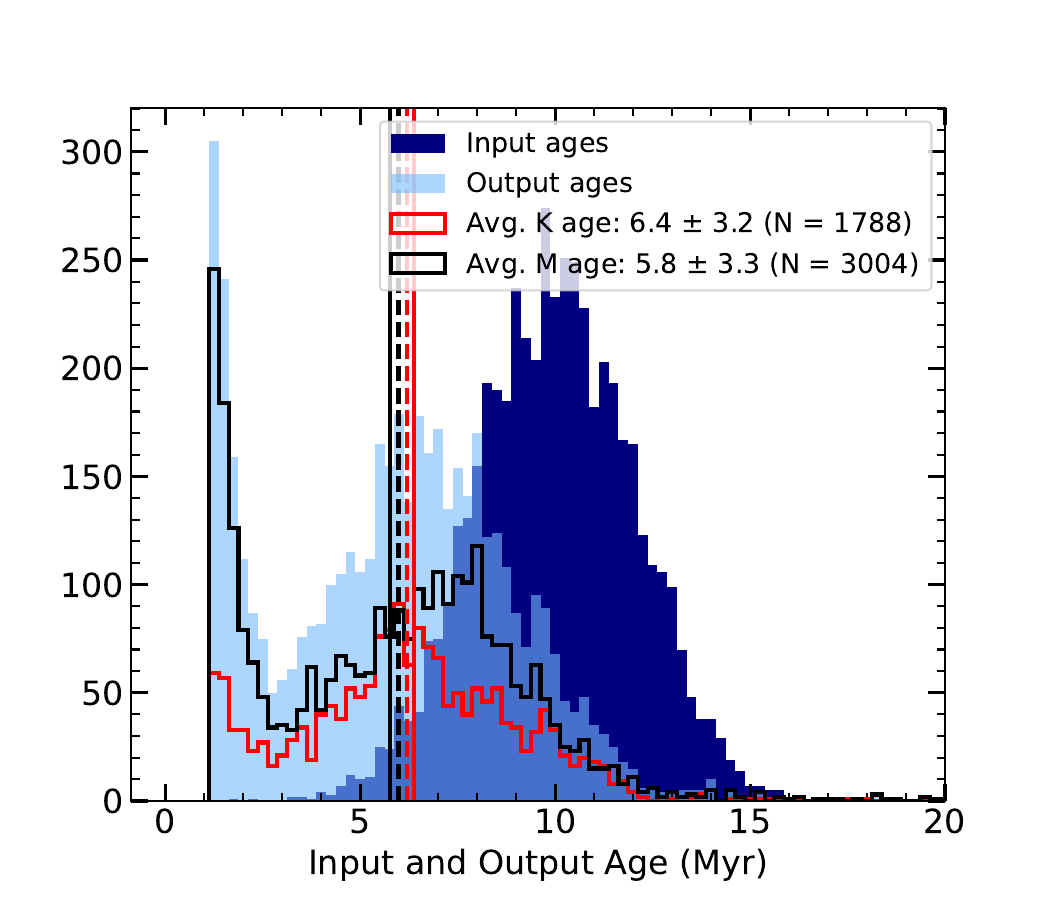}
      \includegraphics[width=0.49\textwidth]{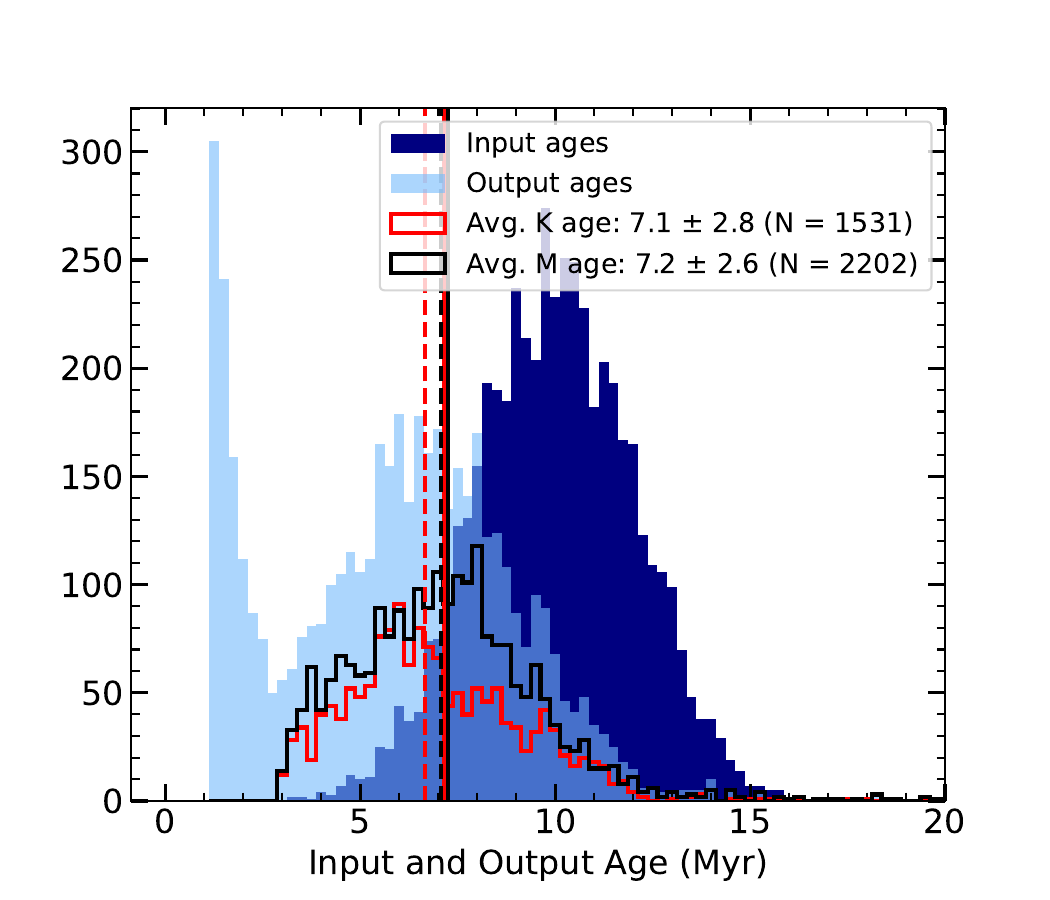}
  \caption {Histograms of input (dark blue) and output (light blue) ages for different simulations. Subsamples of K stars and M stars are plotted as the red and black histograms, respectively. The black (red) solid and dashed lines show the M (K) star mean and median age. (Top:) Single stars with 50\% mean $f_{\rm spot}$ with 20\% intrinsic Gaussian scatter. The mean age and rms scatter for K and M stars are both 7.1 Myr (rms $\sim$ 2) Myr. (Bottom left:) A simulation with binary stars and spots, with 10\% Gaussian scatter added to $f_{spot}$. The average K star age is 6.4 (rms = 3.2) Myr, and the average M star age is 5.8 (rms = 3.3) Myr. The young peak in the output histogram is caused by undetected binary stars and produces a large age spread. (Bottom right:) The same as (b) but with the output age distribution truncated at the local minimum ($\sim$ 3 Myr) between the young peak and the rest of the population. This population with apparent binaries removed has a mean age of 7.1 (rms = 2.8) Myr for K stars, and 7.2 (rms = 2.6) Myr for M stars. The presence of binaries increases the apparent age spread.}
\label{fig:age_hist_all}
\end{figure*}

For each simulation, we inferred an age for each star using the derived \teff\ and luminosity to compare to evolutionary models. In this subsection we present the inferred ages for a series of simulations. Figure \ref{fig:age_hist_all} shows summary histograms for the input and output age distributions of our simulations. We present results from four different simulated scenarios.

Simulation A (top panel of Figure \ref{fig:age_hist_all}) presents a single-star population with spots, but with the bilinearly-interpolated $f_{spot}$ perturbed by a Gaussian with a fractional width of $20\%$. The average K and M ages are both 7.1 (rms $\sim 2$) Myr. Therefore, the spot-only scenario leads to ages that are uniformly younger by $\sim 30$\%, but with age spreads that match the input 2 Myr spread. Thus, starspots alone do not produce an apparent mass-dependent age gradient. The $\sim$ 30\% reduction in age is smaller than the $\sim$ 50\% reduction in age found by \citet{PerezPaolino2024} in the Taurus-Auriga star-forming region ($\tau \sim 1-2$ Myr; \citealt{Krolikowski2021}). This is likely because the spot coverage fraction is age-dependent, and the Taurus targets of \citet{PerezPaolino2024} would thus be more heavily spotted than our simulated Upper Sco population. A larger spot coverage fraction would likely lead to an even larger effect on apparent age for the younger stars.

Simulation B (lower left panel of Figure \ref{fig:age_hist_all}) shows the population of single stars and binaries, with perturbed $f_{spot}$. The binaries appear much younger than the rest of the population, producing a second peak of very young stars. Because the impact of multiplicity is more distinctive for lower-mass stars, the M star population contributes almost all the systems that comprise the young peak, while the K star population appears more spread out with a smaller young peak. The mean ages and spreads reflect the presence of the second peak that is caused by binaries, with the average M star age $\bar{\tau_{M}} = 5.7$ (rms = 3.3) Myr, and the average K star age $\bar{\tau_{K}} = 6.4$ (rms = 3.2) Myr.

It is likely that in a real set of observations, the secondary peak caused by the presence of binaries would be identified as a ``binary main sequence'' and removed from the analysis. Therefore, we also wished to examine the population with binaries included but removing the easily-detectable binary systems (i.e., the young peak of the retrieved-age histogram). The lower right panel of Figure \ref{fig:age_hist_all} shows Simulation B, but with the age distributions truncated at the local minimum for the age distribution of the M stars at $\sim$3 Myr (Simulation C hereafter). In this scenario, mimicking removal of the obvious binaries, the average M star age is $\bar{\tau_{M}} = 7.2$ (rms = 2.6) Myr, and the average K star age $\bar{\tau_{K}} = 7.1$ (rms = 2.8) Myr. 

Because adding scatter to the starspot coverage fraction does not introduce additional scatter into the age measurements (i.e., the rms scatter of Simulation A does not change from the initial input scatter of 2 Myr), any additional scatter beyond the intrinsic 2 Myr scatter is caused by the presence of additional undetected binaries present even in the truncated sample. Although the rms scatter is increased in Simulation C relative to Simulation A, the mean ages for both K and M stars are identical (within error) to Simulation A, which did not include binaries. This suggests that if the majority of binaries can be removed, age measurements using isochrone fitting should retrieve mean or median ages that are predominantly biased by the presence of undetected starspots, rather than binaries, but the age spread will be impacted by undetected multiplicity. 

In our previous paper exploring the impact of binary stars on age measurements \citep{Sullivan2021}, we did not truncate the age distribution at the location of any apparent bimodality in the age distribution, although we did observe a double-peaked age distribution for the M stars. \citet{Sullivan2021} found that in the exact-distance scenario (which we essentially replicated in this work by mimicking {\it Gaia} observations), the average M star age was $\bar{\tau_{M, 2021}} = 7.0$ (rms = 4.3) Myr. The simulation in this paper that most closely matches the analysis of \citet{Sullivan2021} is Simulation C, where the average age for M stars was $\bar{\tau_{M, 2023}} = 7.2$ (rms = 2.6) Myr. In \citet{Sullivan2021} the rms is increased relative to Simulation C. In this work, the starspots drive the single-star age distribution to younger ages, decreasing the total rms scatter and also decreasing the average age of the population in conjunction with the young-star peak caused by the binaries.

The starspots in Simulation B and C are also the cause of the more defined binary sequence in this work relative to \citet{Sullivan2021}. Multiplicity decreases measured \teff\ but increases system luminosity, while starspots decrease both measured \teff\ and luminosity. The combination of these effects causes more separation between the single and binary stars in the simulations of this work than in \citet{Sullivan2021}. This suggests that it may be easier to remove binaries from low-mass stellar populations dominated by M stars than to remove them from even slightly higher-mass samples such as those dominated by K stars. Also, high-mass stellar populations are more likely than low-mass stellar populations to be hosting binaries that are challenging to detect via isochrone fitting, meaning that it will be more difficult to remove binaries from G or K star samples than from M star samples. 

Additional challenges to age measurements will include uncertain distances, which will blur out the bimodality that enables removal of binaries for low-mass stars. Gaia astrometric solutions become worse for young stars because of disks and variability \citep[e.g.,][]{Fitton2022}, extinction, binarity itself \citep{Wood2021}, and more distant stellar populations, meaning that typical parallax errors for young stellar associations may be larger than what we assume in this work. 

\section{Conclusions}
To explore the relative impact of starspots and undetected binary stars on age measurements of young stellar associations, we simulated two spectroscopic surveys of an analog of the Upper Sco association. One simulated survey consisted of only single stars with starspots that were not corrected for, while the other survey was identical but also included undetected binary stars. We found that starspots do not produce an apparent mass-dependent age gradient, but a uniform spot coverage fraction of 50\% leads to an age for both M and K stars that is $\sim$ 30\% younger than the true age, with a slightly larger age spread, assuming the obvious binaries are removed. If the binary sequence is not removed (as in situations with less certain distances), the lower-mass stars appear systematically younger than the higher-mass stars, and both populations appear younger than their true ages. Thus, we found that starspots produce a statistically significant age bias, in keeping with other studies, but found that they cannot replicate the apparent mass-dependent age gradient observed in star-forming regions. Ideally, any census of a young stellar population would identify binaries and either analyze them separately or fully exclude them, but if that is not possible a forward-modeling routine that includes the impacts of both binaries and spots will be important for accurately retrieving mass functions and star formation histories.

\software{astropy \citep{astropy2013, astropy2018, astropy2022}, extinction \citep{Barbary2016}, matplotlib \citep{Hunter2007}, numpy \citep{Harris2020}, scipy \citep{Virtanen2020}}

We thank Michael Gully-Santiago and Ben Tofflemire for helpful feedback on the methods used in this work. We thank the anonymous referee for their comments, which have greatly improved this work. The authors acknowledge the Texas Advanced Computing Center (TACC) at The University of Texas at Austin for providing high-performance computing resources that have contributed to the research results reported within this paper. 

\bibliography{bib}

\begin{thebibliography}{}
\expandafter\ifx\csname natexlab\endcsname\relax\def\natexlab#1{#1}\fi
\providecommand{\url}[1]{\href{#1}{#1}}
\providecommand{\dodoi}[1]{doi:~\href{http://doi.org/#1}{\nolinkurl{#1}}}
\providecommand{\doeprint}[1]{\href{http://ascl.net/#1}{\nolinkurl{http://ascl.net/#1}}}
\providecommand{\doarXiv}[1]{\href{https://arxiv.org/abs/#1}{\nolinkurl{https://arxiv.org/abs/#1}}}

\bibitem[{{Astropy Collaboration} {et~al.}(2013){Astropy Collaboration},
  {Robitaille}, {Tollerud}, {Greenfield}, {Droettboom}, {Bray}, {Aldcroft},
  {Davis}, {Ginsburg}, {Price-Whelan}, {Kerzendorf}, {Conley}, {Crighton},
  {Barbary}, {Muna}, {Ferguson}, {Grollier}, {Parikh}, {Nair}, {Unther},
  {Deil}, {Woillez}, {Conseil}, {Kramer}, {Turner}, {Singer}, {Fox}, {Weaver},
  {Zabalza}, {Edwards}, {Azalee Bostroem}, {Burke}, {Casey}, {Crawford},
  {Dencheva}, {Ely}, {Jenness}, {Labrie}, {Lim}, {Pierfederici}, {Pontzen},
  {Ptak}, {Refsdal}, {Servillat}, \& {Streicher}}]{astropy2013}
{Astropy Collaboration}, {Robitaille}, T.~P., {Tollerud}, E.~J., {et~al.} 2013,
  \aap, 558, A33, \dodoi{10.1051/0004-6361/201322068}

\bibitem[{{Astropy Collaboration} {et~al.}(2018){Astropy Collaboration},
  {Price-Whelan}, {Sip{\H{o}}cz}, {G{\"u}nther}, {Lim}, {Crawford}, {Conseil},
  {Shupe}, {Craig}, {Dencheva}, {Ginsburg}, {Vand erPlas}, {Bradley},
  {P{\'e}rez-Su{\'a}rez}, {de Val-Borro}, {Aldcroft}, {Cruz}, {Robitaille},
  {Tollerud}, {Ardelean}, {Babej}, {Bach}, {Bachetti}, {Bakanov}, {Bamford},
  {Barentsen}, {Barmby}, {Baumbach}, {Berry}, {Biscani}, {Boquien}, {Bostroem},
  {Bouma}, {Brammer}, {Bray}, {Breytenbach}, {Buddelmeijer}, {Burke},
  {Calderone}, {Cano Rodr{\'\i}guez}, {Cara}, {Cardoso}, {Cheedella}, {Copin},
  {Corrales}, {Crichton}, {D'Avella}, {Deil}, {Depagne}, {Dietrich}, {Donath},
  {Droettboom}, {Earl}, {Erben}, {Fabbro}, {Ferreira}, {Finethy}, {Fox},
  {Garrison}, {Gibbons}, {Goldstein}, {Gommers}, {Greco}, {Greenfield},
  {Groener}, {Grollier}, {Hagen}, {Hirst}, {Homeier}, {Horton}, {Hosseinzadeh},
  {Hu}, {Hunkeler}, {Ivezi{\'c}}, {Jain}, {Jenness}, {Kanarek}, {Kendrew},
  {Kern}, {Kerzendorf}, {Khvalko}, {King}, {Kirkby}, {Kulkarni}, {Kumar},
  {Lee}, {Lenz}, {Littlefair}, {Ma}, {Macleod}, {Mastropietro}, {McCully},
  {Montagnac}, {Morris}, {Mueller}, {Mumford}, {Muna}, {Murphy}, {Nelson},
  {Nguyen}, {Ninan}, {N{\"o}the}, {Ogaz}, {Oh}, {Parejko}, {Parley}, {Pascual},
  {Patil}, {Patil}, {Plunkett}, {Prochaska}, {Rastogi}, {Reddy Janga},
  {Sabater}, {Sakurikar}, {Seifert}, {Sherbert}, {Sherwood-Taylor}, {Shih},
  {Sick}, {Silbiger}, {Singanamalla}, {Singer}, {Sladen}, {Sooley},
  {Sornarajah}, {Streicher}, {Teuben}, {Thomas}, {Tremblay}, {Turner},
  {Terr{\'o}n}, {van Kerkwijk}, {de la Vega}, {Watkins}, {Weaver}, {Whitmore},
  {Woillez}, {Zabalza}, \& {Astropy Contributors}}]{astropy2018}
{Astropy Collaboration}, {Price-Whelan}, A.~M., {Sip{\H{o}}cz}, B.~M., {et~al.}
  2018, \aj, 156, 123, \dodoi{10.3847/1538-3881/aabc4f}

\bibitem[{{Astropy Collaboration} {et~al.}(2022){Astropy Collaboration},
  {Price-Whelan}, {Lim}, {Earl}, {Starkman}, {Bradley}, {Shupe}, {Patil},
  {Corrales}, {Brasseur}, {N{\"o}the}, {Donath}, {Tollerud}, {Morris},
  {Ginsburg}, {Vaher}, {Weaver}, {Tocknell}, {Jamieson}, {van Kerkwijk},
  {Robitaille}, {Merry}, {Bachetti}, {G{\"u}nther}, {Aldcroft},
  {Alvarado-Montes}, {Archibald}, {B{\'o}di}, {Bapat}, {Barentsen},
  {Baz{\'a}n}, {Biswas}, {Boquien}, {Burke}, {Cara}, {Cara}, {Conroy},
  {Conseil}, {Craig}, {Cross}, {Cruz}, {D'Eugenio}, {Dencheva}, {Devillepoix},
  {Dietrich}, {Eigenbrot}, {Erben}, {Ferreira}, {Foreman-Mackey}, {Fox},
  {Freij}, {Garg}, {Geda}, {Glattly}, {Gondhalekar}, {Gordon}, {Grant},
  {Greenfield}, {Groener}, {Guest}, {Gurovich}, {Handberg}, {Hart},
  {Hatfield-Dodds}, {Homeier}, {Hosseinzadeh}, {Jenness}, {Jones}, {Joseph},
  {Kalmbach}, {Karamehmetoglu}, {Ka{\l}uszy{\'n}ski}, {Kelley}, {Kern},
  {Kerzendorf}, {Koch}, {Kulumani}, {Lee}, {Ly}, {Ma}, {MacBride}, {Maljaars},
  {Muna}, {Murphy}, {Norman}, {O'Steen}, {Oman}, {Pacifici}, {Pascual},
  {Pascual-Granado}, {Patil}, {Perren}, {Pickering}, {Rastogi}, {Roulston},
  {Ryan}, {Rykoff}, {Sabater}, {Sakurikar}, {Salgado}, {Sanghi}, {Saunders},
  {Savchenko}, {Schwardt}, {Seifert-Eckert}, {Shih}, {Jain}, {Shukla}, {Sick},
  {Simpson}, {Singanamalla}, {Singer}, {Singhal}, {Sinha}, {Sip{\H{o}}cz},
  {Spitler}, {Stansby}, {Streicher}, {{\v{S}}umak}, {Swinbank}, {Taranu},
  {Tewary}, {Tremblay}, {Val-Borro}, {Van Kooten}, {Vasovi{\'c}}, {Verma}, {de
  Miranda Cardoso}, {Williams}, {Wilson}, {Winkel}, {Wood-Vasey}, {Xue},
  {Yoachim}, {Zhang}, {Zonca}, \& {Astropy Project Contributors}}]{astropy2022}
{Astropy Collaboration}, {Price-Whelan}, A.~M., {Lim}, P.~L., {et~al.} 2022,
  \apj, 935, 167, \dodoi{10.3847/1538-4357/ac7c74}

\bibitem[{Barbary(2016)}]{Barbary2016}
Barbary, K. 2016, extinction v0.3.0,  Zenodo, \dodoi{10.5281/zenodo.804967}

\bibitem[{{Cao} \& {Pinsonneault}(2022)}]{Cao2022b}
{Cao}, L., \& {Pinsonneault}, M.~H. 2022, \mnras, 517, 2165,
  \dodoi{10.1093/mnras/stac2706}

\bibitem[{{Cao} {et~al.}(2022){Cao}, {Pinsonneault}, {Hillenbrand}, \&
  {Kuhn}}]{Cao2022a}
{Cao}, L., {Pinsonneault}, M.~H., {Hillenbrand}, L.~A., \& {Kuhn}, M.~A. 2022,
  \apj, 924, 84, \dodoi{10.3847/1538-4357/ac307f}

\bibitem[{{Cardelli} {et~al.}(1989){Cardelli}, {Clayton}, \&
  {Mathis}}]{Cardelli1989}
{Cardelli}, J.~A., {Clayton}, G.~C., \& {Mathis}, J.~S. 1989, \apj, 345, 245,
  \dodoi{10.1086/167900}

\bibitem[{{Chabrier}(2003)}]{Chabrier2003}
{Chabrier}, G. 2003, \pasp, 115, 763, \dodoi{10.1086/376392}

\bibitem[{{Chabrier}(2005)}]{Chabrier2005}
{Chabrier}, G. 2005, in Astrophysics and Space Science Library, Vol. 327, The
  Initial Mass Function 50 Years Later, ed. E.~{Corbelli}, F.~{Palla}, \&
  H.~{Zinnecker}, 41, \dodoi{10.1007/978-1-4020-3407-7_5}

\bibitem[{{Choi} {et~al.}(2016){Choi}, {Dotter}, {Conroy}, {Cantiello},
  {Paxton}, \& {Johnson}}]{Choi2016}
{Choi}, J., {Dotter}, A., {Conroy}, C., {et~al.} 2016, \apj, 823, 102,
  \dodoi{10.3847/0004-637X/823/2/102}

\bibitem[{{De Rosa} {et~al.}(2014){De Rosa}, {Patience}, {Wilson}, {Schneider},
  {Wiktorowicz}, {Vigan}, {Marois}, {Song}, {Macintosh}, {Graham}, {Doyon},
  {Bessell}, {Thomas}, \& {Lai}}]{DeRosa2014}
{De Rosa}, R.~J., {Patience}, J., {Wilson}, P.~A., {et~al.} 2014, \mnras, 437,
  1216, \dodoi{10.1093/mnras/stt1932}

\bibitem[{{Dotter}(2016)}]{Dotter2016}
{Dotter}, A. 2016, \apjs, 222, 8, \dodoi{10.3847/0067-0049/222/1/8}

\bibitem[{{Duch{\^e}ne} \& {Kraus}(2013)}]{Duchene2013}
{Duch{\^e}ne}, G., \& {Kraus}, A. 2013, \araa, 51, 269,
  \dodoi{10.1146/annurev-astro-081710-102602}

\bibitem[{{Elmegreen} \& {Lada}(1977)}]{Elmegreen1977}
{Elmegreen}, B.~G., \& {Lada}, C.~J. 1977, \apj, 214, 725,
  \dodoi{10.1086/155302}

\bibitem[{{Feiden}(2016)}]{Feiden2016}
{Feiden}, G.~A. 2016, \aap, 593, A99, \dodoi{10.1051/0004-6361/201527613}

\bibitem[{{Fitton} {et~al.}(2022){Fitton}, {Tofflemire}, \&
  {Kraus}}]{Fitton2022}
{Fitton}, S., {Tofflemire}, B.~M., \& {Kraus}, A.~L. 2022, Research Notes of
  the American Astronomical Society, 6, 18, \dodoi{10.3847/2515-5172/ac4bb7}

\bibitem[{{Gaia Collaboration} {et~al.}(2016){Gaia Collaboration}, {Prusti},
  {de Bruijne}, {Brown}, {Vallenari}, {Babusiaux}, {Bailer-Jones}, {Bastian},
  {Biermann}, {Evans}, {Eyer}, {Jansen}, {Jordi}, {Klioner}, {Lammers},
  {Lindegren}, {Luri}, {Mignard}, {Milligan}, {Panem}, {Poinsignon},
  {Pourbaix}, {Randich}, {Sarri}, {Sartoretti}, {Siddiqui}, {Soubiran},
  {Valette}, {van Leeuwen}, {Walton}, {Aerts}, {Arenou}, {Cropper}, {Drimmel},
  {H{\o}g}, {Katz}, {Lattanzi}, {O'Mullane}, {Grebel}, {Holland}, {Huc},
  {Passot}, {Bramante}, {Cacciari}, {Casta{\~n}eda}, {Chaoul}, {Cheek}, {De
  Angeli}, {Fabricius}, {Guerra}, {Hern{\'a}ndez}, {Jean-Antoine-Piccolo},
  {Masana}, {Messineo}, {Mowlavi}, {Nienartowicz}, {Ord{\'o}{\~n}ez-Blanco},
  {Panuzzo}, {Portell}, {Richards}, {Riello}, {Seabroke}, {Tanga},
  {Th{\'e}venin}, {Torra}, {Els}, {Gracia-Abril}, {Comoretto},
  {Garcia-Reinaldos}, {Lock}, {Mercier}, {Altmann}, {Andrae}, {Astraatmadja},
  {Bellas-Velidis}, {Benson}, {Berthier}, {Blomme}, {Busso}, {Carry},
  {Cellino}, {Clementini}, {Cowell}, {Creevey}, {Cuypers}, {Davidson}, {De
  Ridder}, {de Torres}, {Delchambre}, {Dell'Oro}, {Ducourant}, {Fr{\'e}mat},
  {Garc{\'\i}a-Torres}, {Gosset}, {Halbwachs}, {Hambly}, {Harrison}, {Hauser},
  {Hestroffer}, {Hodgkin}, {Huckle}, {Hutton}, {Jasniewicz}, {Jordan},
  {Kontizas}, {Korn}, {Lanzafame}, {Manteiga}, {Moitinho}, {Muinonen},
  {Osinde}, {Pancino}, {Pauwels}, {Petit}, {Recio-Blanco}, {Robin}, {Sarro},
  {Siopis}, {Smith}, {Smith}, {Sozzetti}, {Thuillot}, {van Reeven}, {Viala},
  {Abbas}, {Abreu Aramburu}, {Accart}, {Aguado}, {Allan}, {Allasia},
  {Altavilla}, {{\'A}lvarez}, {Alves}, {Anderson}, {Andrei}, {Anglada Varela},
  {Antiche}, {Antoja}, {Ant{\'o}n}, {Arcay}, {Atzei}, {Ayache}, {Bach},
  {Baker}, {Balaguer-N{\'u}{\~n}ez}, {Barache}, {Barata}, {Barbier}, {Barblan},
  {Baroni}, {Barrado y Navascu{\'e}s}, {Barros}, {Barstow}, {Becciani},
  {Bellazzini}, {Bellei}, {Bello Garc{\'\i}a}, {Belokurov}, {Bendjoya},
  {Berihuete}, {Bianchi}, {Bienaym{\'e}}, {Billebaud}, {Blagorodnova},
  {Blanco-Cuaresma}, {Boch}, {Bombrun}, {Borrachero}, {Bouquillon}, {Bourda},
  {Bouy}, {Bragaglia}, {Breddels}, {Brouillet}, {Br{\"u}semeister},
  {Bucciarelli}, {Budnik}, {Burgess}, {Burgon}, {Burlacu}, {Busonero}, {Buzzi},
  {Caffau}, {Cambras}, {Campbell}, {Cancelliere}, {Cantat-Gaudin}, {Carlucci},
  {Carrasco}, {Castellani}, {Charlot}, {Charnas}, {Charvet}, {Chassat},
  {Chiavassa}, {Clotet}, {Cocozza}, {Collins}, {Collins}, {Costigan}, {Crifo},
  {Cross}, {Crosta}, {Crowley}, {Dafonte}, {Damerdji}, {Dapergolas}, {David},
  {David}, {De Cat}, {de Felice}, {de Laverny}, {De Luise}, {De March}, {de
  Martino}, {de Souza}, {Debosscher}, {del Pozo}, {Delbo}, {Delgado},
  {Delgado}, {di Marco}, {Di Matteo}, {Diakite}, {Distefano}, {Dolding}, {Dos
  Anjos}, {Drazinos}, {Dur{\'a}n}, {Dzigan}, {Ecale}, {Edvardsson}, {Enke},
  {Erdmann}, {Escolar}, {Espina}, {Evans}, {Eynard Bontemps}, {Fabre},
  {Fabrizio}, {Faigler}, {Falc{\~a}o}, {Farr{\`a}s Casas}, {Faye}, {Federici},
  {Fedorets}, {Fern{\'a}ndez-Hern{\'a}ndez}, {Fernique}, {Fienga}, {Figueras},
  {Filippi}, {Findeisen}, {Fonti}, {Fouesneau}, {Fraile}, {Fraser}, {Fuchs},
  {Furnell}, {Gai}, {Galleti}, {Galluccio}, {Garabato}, {Garc{\'\i}a-Sedano},
  {Gar{\'e}}, {Garofalo}, {Garralda}, {Gavras}, {Gerssen}, {Geyer}, {Gilmore},
  {Girona}, {Giuffrida}, {Gomes}, {Gonz{\'a}lez-Marcos},
  {Gonz{\'a}lez-N{\'u}{\~n}ez}, {Gonz{\'a}lez-Vidal}, {Granvik}, {Guerrier},
  {Guillout}, {Guiraud}, {G{\'u}rpide}, {Guti{\'e}rrez-S{\'a}nchez}, {Guy},
  {Haigron}, {Hatzidimitriou}, {Haywood}, {Heiter}, {Helmi}, {Hobbs},
  {Hofmann}, {Holl}, {Holland}, {Hunt}, {Hypki}, {Icardi}, {Irwin}, {Jevardat
  de Fombelle}, {Jofr{\'e}}, {Jonker}, {Jorissen}, {Julbe}, {Karampelas},
  {Kochoska}, {Kohley}, {Kolenberg}, {Kontizas}, {Koposov}, {Kordopatis},
  {Koubsky}, {Kowalczyk}, {Krone-Martins}, {Kudryashova}, {Kull}, {Bachchan},
  {Lacoste-Seris}, {Lanza}, {Lavigne}, {Le Poncin-Lafitte}, {Lebreton},
  {Lebzelter}, {Leccia}, {Leclerc}, {Lecoeur-Taibi}, {Lemaitre}, {Lenhardt},
  {Leroux}, {Liao}, {Licata}, {Lindstr{\o}m}, {Lister}, {Livanou}, {Lobel},
  {L{\"o}ffler}, {L{\'o}pez}, {Lopez-Lozano}, {Lorenz}, {Loureiro},
  {MacDonald}, {Magalh{\~a}es Fernandes}, {Managau}, {Mann}, {Mantelet},
  {Marchal}, {Marchant}, {Marconi}, {Marie}, {Marinoni}, {Marrese},
  {Marschalk{\'o}}, {Marshall}, {Mart{\'\i}n-Fleitas}, {Martino}, {Mary},
  {Matijevi{\v{c}}}, {Mazeh}, {McMillan}, {Messina}, {Mestre}, {Michalik},
  {Millar}, {Miranda}, {Molina}, {Molinaro}, {Molinaro}, {Moln{\'a}r},
  {Moniez}, {Montegriffo}, {Monteiro}, {Mor}, {Mora}, {Morbidelli}, {Morel},
  {Morgenthaler}, {Morley}, {Morris}, {Mulone}, {Muraveva}, {Musella},
  {Narbonne}, {Nelemans}, {Nicastro}, {Noval}, {Ord{\'e}novic},
  {Ordieres-Mer{\'e}}, {Osborne}, {Pagani}, {Pagano}, {Pailler}, {Palacin},
  {Palaversa}, {Parsons}, {Paulsen}, {Pecoraro}, {Pedrosa}, {Pentik{\"a}inen},
  {Pereira}, {Pichon}, {Piersimoni}, {Pineau}, {Plachy}, {Plum}, {Poujoulet},
  {Pr{\v{s}}a}, {Pulone}, {Ragaini}, {Rago}, {Rambaux}, {Ramos-Lerate},
  {Ranalli}, {Rauw}, {Read}, {Regibo}, {Renk}, {Reyl{\'e}}, {Ribeiro},
  {Rimoldini}, {Ripepi}, {Riva}, {Rixon}, {Roelens}, {Romero-G{\'o}mez},
  {Rowell}, {Royer}, {Rudolph}, {Ruiz-Dern}, {Sadowski}, {Sagrist{\`a}
  Sell{\'e}s}, {Sahlmann}, {Salgado}, {Salguero}, {Sarasso}, {Savietto},
  {Schnorhk}, {Schultheis}, {Sciacca}, {Segol}, {Segovia}, {Segransan},
  {Serpell}, {Shih}, {Smareglia}, {Smart}, {Smith}, {Solano}, {Solitro},
  {Sordo}, {Soria Nieto}, {Souchay}, {Spagna}, {Spoto}, {Stampa}, {Steele},
  {Steidelm{\"u}ller}, {Stephenson}, {Stoev}, {Suess}, {S{\"u}veges}, {Surdej},
  {Szabados}, {Szegedi-Elek}, {Tapiador}, {Taris}, {Tauran}, {Taylor},
  {Teixeira}, {Terrett}, {Tingley}, {Trager}, {Turon}, {Ulla}, {Utrilla},
  {Valentini}, {van Elteren}, {Van Hemelryck}, {van Leeuwen}, {Varadi},
  {Vecchiato}, {Veljanoski}, {Via}, {Vicente}, {Vogt}, {Voss}, {Votruba},
  {Voutsinas}, {Walmsley}, {Weiler}, {Weingrill}, {Werner}, {Wevers},
  {Whitehead}, {Wyrzykowski}, {Yoldas}, {{\v{Z}}erjal}, {Zucker}, {Zurbach},
  {Zwitter}, {Alecu}, {Allen}, {Allende Prieto}, {Amorim},
  {Anglada-Escud{\'e}}, {Arsenijevic}, {Azaz}, {Balm}, {Beck}, {Bernstein},
  {Bigot}, {Bijaoui}, {Blasco}, {Bonfigli}, {Bono}, {Boudreault}, {Bressan},
  {Brown}, {Brunet}, {Bunclark}, {Buonanno}, {Butkevich}, {Carret}, {Carrion},
  {Chemin}, {Ch{\'e}reau}, {Corcione}, {Darmigny}, {de Boer}, {de Teodoro}, {de
  Zeeuw}, {Delle Luche}, {Domingues}, {Dubath}, {Fodor}, {Fr{\'e}zouls},
  {Fries}, {Fustes}, {Fyfe}, {Gallardo}, {Gallegos}, {Gardiol}, {Gebran},
  {Gomboc}, {G{\'o}mez}, {Grux}, {Gueguen}, {Heyrovsky}, {Hoar}, {Iannicola},
  {Isasi Parache}, {Janotto}, {Joliet}, {Jonckheere}, {Keil}, {Kim},
  {Klagyivik}, {Klar}, {Knude}, {Kochukhov}, {Kolka}, {Kos}, {Kutka}, {Lainey},
  {LeBouquin}, {Liu}, {Loreggia}, {Makarov}, {Marseille}, {Martayan},
  {Martinez-Rubi}, {Massart}, {Meynadier}, {Mignot}, {Munari}, {Nguyen},
  {Nordlander}, {Ocvirk}, {O'Flaherty}, {Olias Sanz}, {Ortiz}, {Osorio},
  {Oszkiewicz}, {Ouzounis}, {Palmer}, {Park}, {Pasquato}, {Peltzer}, {Peralta},
  {P{\'e}turaud}, {Pieniluoma}, {Pigozzi}, {Poels}, {Prat}, {Prod'homme},
  {Raison}, {Rebordao}, {Risquez}, {Rocca-Volmerange}, {Rosen}, {Ruiz-Fuertes},
  {Russo}, {Sembay}, {Serraller Vizcaino}, {Short}, {Siebert}, {Silva},
  {Sinachopoulos}, {Slezak}, {Soffel}, {Sosnowska}, {Strai{\v{z}}ys}, {ter
  Linden}, {Terrell}, {Theil}, {Tiede}, {Troisi}, {Tsalmantza}, {Tur},
  {Vaccari}, {Vachier}, {Valles}, {Van Hamme}, {Veltz}, {Virtanen}, {Wallut},
  {Wichmann}, {Wilkinson}, {Ziaeepour}, \& {Zschocke}}]{Gaia2016}
{Gaia Collaboration}, {Prusti}, T., {de Bruijne}, J.~H.~J., {et~al.} 2016,
  \aap, 595, A1, \dodoi{10.1051/0004-6361/201629272}

\bibitem[{{Gaia Collaboration} {et~al.}(2018){Gaia Collaboration}, {Brown},
  {Vallenari}, {Prusti}, {de Bruijne}, {Babusiaux}, {Bailer-Jones}, {Biermann},
  {Evans}, {Eyer}, {Jansen}, {Jordi}, {Klioner}, {Lammers}, {Lindegren},
  {Luri}, {Mignard}, {Panem}, {Pourbaix}, {Randich}, {Sartoretti}, {Siddiqui},
  {Soubiran}, {van Leeuwen}, {Walton}, {Arenou}, {Bastian}, {Cropper},
  {Drimmel}, {Katz}, {Lattanzi}, {Bakker}, {Cacciari}, {Casta{\~n}eda},
  {Chaoul}, {Cheek}, {De Angeli}, {Fabricius}, {Guerra}, {Holl}, {Masana},
  {Messineo}, {Mowlavi}, {Nienartowicz}, {Panuzzo}, {Portell}, {Riello},
  {Seabroke}, {Tanga}, {Th{\'e}venin}, {Gracia-Abril}, {Comoretto},
  {Garcia-Reinaldos}, {Teyssier}, {Altmann}, {Andrae}, {Audard},
  {Bellas-Velidis}, {Benson}, {Berthier}, {Blomme}, {Burgess}, {Busso},
  {Carry}, {Cellino}, {Clementini}, {Clotet}, {Creevey}, {Davidson}, {De
  Ridder}, {Delchambre}, {Dell'Oro}, {Ducourant},
  {Fern{\'a}ndez-Hern{\'a}ndez}, {Fouesneau}, {Fr{\'e}mat}, {Galluccio},
  {Garc{\'\i}a-Torres}, {Gonz{\'a}lez-N{\'u}{\~n}ez}, {Gonz{\'a}lez-Vidal},
  {Gosset}, {Guy}, {Halbwachs}, {Hambly}, {Harrison}, {Hern{\'a}ndez},
  {Hestroffer}, {Hodgkin}, {Hutton}, {Jasniewicz}, {Jean-Antoine-Piccolo},
  {Jordan}, {Korn}, {Krone-Martins}, {Lanzafame}, {Lebzelter}, {L{\"o}ffler},
  {Manteiga}, {Marrese}, {Mart{\'\i}n-Fleitas}, {Moitinho}, {Mora}, {Muinonen},
  {Osinde}, {Pancino}, {Pauwels}, {Petit}, {Recio-Blanco}, {Richards},
  {Rimoldini}, {Robin}, {Sarro}, {Siopis}, {Smith}, {Sozzetti}, {S{\"u}veges},
  {Torra}, {van Reeven}, {Abbas}, {Abreu Aramburu}, {Accart}, {Aerts},
  {Altavilla}, {{\'A}lvarez}, {Alvarez}, {Alves}, {Anderson}, {Andrei},
  {Anglada Varela}, {Antiche}, {Antoja}, {Arcay}, {Astraatmadja}, {Bach},
  {Baker}, {Balaguer-N{\'u}{\~n}ez}, {Balm}, {Barache}, {Barata}, {Barbato},
  {Barblan}, {Barklem}, {Barrado}, {Barros}, {Barstow}, {Bartholom{\'e}
  Mu{\~n}oz}, {Bassilana}, {Becciani}, {Bellazzini}, {Berihuete}, {Bertone},
  {Bianchi}, {Bienaym{\'e}}, {Blanco-Cuaresma}, {Boch}, {Boeche}, {Bombrun},
  {Borrachero}, {Bossini}, {Bouquillon}, {Bourda}, {Bragaglia}, {Bramante},
  {Breddels}, {Bressan}, {Brouillet}, {Br{\"u}semeister}, {Brugaletta},
  {Bucciarelli}, {Burlacu}, {Busonero}, {Butkevich}, {Buzzi}, {Caffau},
  {Cancelliere}, {Cannizzaro}, {Cantat-Gaudin}, {Carballo}, {Carlucci},
  {Carrasco}, {Casamiquela}, {Castellani}, {Castro-Ginard}, {Charlot},
  {Chemin}, {Chiavassa}, {Cocozza}, {Costigan}, {Cowell}, {Crifo}, {Crosta},
  {Crowley}, {Cuypers}, {Dafonte}, {Damerdji}, {Dapergolas}, {David}, {David},
  {de Laverny}, {De Luise}, {De March}, {de Martino}, {de Souza}, {de Torres},
  {Debosscher}, {del Pozo}, {Delbo}, {Delgado}, {Delgado}, {Di Matteo},
  {Diakite}, {Diener}, {Distefano}, {Dolding}, {Drazinos}, {Dur{\'a}n},
  {Edvardsson}, {Enke}, {Eriksson}, {Esquej}, {Eynard Bontemps}, {Fabre},
  {Fabrizio}, {Faigler}, {Falc{\~a}o}, {Farr{\`a}s Casas}, {Federici},
  {Fedorets}, {Fernique}, {Figueras}, {Filippi}, {Findeisen}, {Fonti},
  {Fraile}, {Fraser}, {Fr{\'e}zouls}, {Gai}, {Galleti}, {Garabato},
  {Garc{\'\i}a-Sedano}, {Garofalo}, {Garralda}, {Gavel}, {Gavras}, {Gerssen},
  {Geyer}, {Giacobbe}, {Gilmore}, {Girona}, {Giuffrida}, {Glass}, {Gomes},
  {Granvik}, {Gueguen}, {Guerrier}, {Guiraud}, {Guti{\'e}rrez-S{\'a}nchez},
  {Haigron}, {Hatzidimitriou}, {Hauser}, {Haywood}, {Heiter}, {Helmi}, {Heu},
  {Hilger}, {Hobbs}, {Hofmann}, {Holland}, {Huckle}, {Hypki}, {Icardi},
  {Jan{\ss}en}, {Jevardat de Fombelle}, {Jonker}, {Juh{\'a}sz}, {Julbe},
  {Karampelas}, {Kewley}, {Klar}, {Kochoska}, {Kohley}, {Kolenberg},
  {Kontizas}, {Kontizas}, {Koposov}, {Kordopatis}, {Kostrzewa-Rutkowska},
  {Koubsky}, {Lambert}, {Lanza}, {Lasne}, {Lavigne}, {Le Fustec}, {Le
  Poncin-Lafitte}, {Lebreton}, {Leccia}, {Leclerc}, {Lecoeur-Taibi},
  {Lenhardt}, {Leroux}, {Liao}, {Licata}, {Lindstr{\o}m}, {Lister}, {Livanou},
  {Lobel}, {L{\'o}pez}, {Managau}, {Mann}, {Mantelet}, {Marchal}, {Marchant},
  {Marconi}, {Marinoni}, {Marschalk{\'o}}, {Marshall}, {Martino}, {Marton},
  {Mary}, {Massari}, {Matijevi{\v{c}}}, {Mazeh}, {McMillan}, {Messina},
  {Michalik}, {Millar}, {Molina}, {Molinaro}, {Moln{\'a}r}, {Montegriffo},
  {Mor}, {Morbidelli}, {Morel}, {Morris}, {Mulone}, {Muraveva}, {Musella},
  {Nelemans}, {Nicastro}, {Noval}, {O'Mullane}, {Ord{\'e}novic},
  {Ord{\'o}{\~n}ez-Blanco}, {Osborne}, {Pagani}, {Pagano}, {Pailler},
  {Palacin}, {Palaversa}, {Panahi}, {Pawlak}, {Piersimoni}, {Pineau}, {Plachy},
  {Plum}, {Poggio}, {Poujoulet}, {Pr{\v{s}}a}, {Pulone}, {Racero}, {Ragaini},
  {Rambaux}, {Ramos-Lerate}, {Regibo}, {Reyl{\'e}}, {Riclet}, {Ripepi}, {Riva},
  {Rivard}, {Rixon}, {Roegiers}, {Roelens}, {Romero-G{\'o}mez}, {Rowell},
  {Royer}, {Ruiz-Dern}, {Sadowski}, {Sagrist{\`a} Sell{\'e}s}, {Sahlmann},
  {Salgado}, {Salguero}, {Sanna}, {Santana-Ros}, {Sarasso}, {Savietto},
  {Schultheis}, {Sciacca}, {Segol}, {Segovia}, {S{\'e}gransan}, {Shih},
  {Siltala}, {Silva}, {Smart}, {Smith}, {Solano}, {Solitro}, {Sordo}, {Soria
  Nieto}, {Souchay}, {Spagna}, {Spoto}, {Stampa}, {Steele},
  {Steidelm{\"u}ller}, {Stephenson}, {Stoev}, {Suess}, {Surdej}, {Szabados},
  {Szegedi-Elek}, {Tapiador}, {Taris}, {Tauran}, {Taylor}, {Teixeira},
  {Terrett}, {Teyssandier}, {Thuillot}, {Titarenko}, {Torra Clotet}, {Turon},
  {Ulla}, {Utrilla}, {Uzzi}, {Vaillant}, {Valentini}, {Valette}, {van Elteren},
  {Van Hemelryck}, {van Leeuwen}, {Vaschetto}, {Vecchiato}, {Veljanoski},
  {Viala}, {Vicente}, {Vogt}, {von Essen}, {Voss}, {Votruba}, {Voutsinas},
  {Walmsley}, {Weiler}, {Wertz}, {Wevers}, {Wyrzykowski}, {Yoldas},
  {{\v{Z}}erjal}, {Ziaeepour}, {Zorec}, {Zschocke}, {Zucker}, {Zurbach}, \&
  {Zwitter}}]{Gaia2018}
{Gaia Collaboration}, {Brown}, A.~G.~A., {Vallenari}, A., {et~al.} 2018, \aap,
  616, A1, \dodoi{10.1051/0004-6361/201833051}

\bibitem[{{Gaia Collaboration} {et~al.}(2021){Gaia Collaboration}, {Brown},
  {Vallenari}, {Prusti}, {de Bruijne}, {Babusiaux}, {Biermann}, {Creevey},
  {Evans}, {Eyer}, {Hutton}, {Jansen}, {Jordi}, {Klioner}, {Lammers},
  {Lindegren}, {Luri}, {Mignard}, {Panem}, {Pourbaix}, {Randich}, {Sartoretti},
  {Soubiran}, {Walton}, {Arenou}, {Bailer-Jones}, {Bastian}, {Cropper},
  {Drimmel}, {Katz}, {Lattanzi}, {van Leeuwen}, {Bakker}, {Cacciari},
  {Casta{\~n}eda}, {De Angeli}, {Ducourant}, {Fabricius}, {Fouesneau},
  {Fr{\'e}mat}, {Guerra}, {Guerrier}, {Guiraud}, {Jean-Antoine Piccolo},
  {Masana}, {Messineo}, {Mowlavi}, {Nicolas}, {Nienartowicz}, {Pailler},
  {Panuzzo}, {Riclet}, {Roux}, {Seabroke}, {Sordo}, {Tanga}, {Th{\'e}venin},
  {Gracia-Abril}, {Portell}, {Teyssier}, {Altmann}, {Andrae}, {Bellas-Velidis},
  {Benson}, {Berthier}, {Blomme}, {Brugaletta}, {Burgess}, {Busso}, {Carry},
  {Cellino}, {Cheek}, {Clementini}, {Damerdji}, {Davidson}, {Delchambre},
  {Dell'Oro}, {Fern{\'a}ndez-Hern{\'a}ndez}, {Galluccio}, {Garc{\'\i}a-Lario},
  {Garcia-Reinaldos}, {Gonz{\'a}lez-N{\'u}{\~n}ez}, {Gosset}, {Haigron},
  {Halbwachs}, {Hambly}, {Harrison}, {Hatzidimitriou}, {Heiter},
  {Hern{\'a}ndez}, {Hestroffer}, {Hodgkin}, {Holl}, {Jan{\ss}en}, {Jevardat de
  Fombelle}, {Jordan}, {Krone-Martins}, {Lanzafame}, {L{\"o}ffler}, {Lorca},
  {Manteiga}, {Marchal}, {Marrese}, {Moitinho}, {Mora}, {Muinonen}, {Osborne},
  {Pancino}, {Pauwels}, {Petit}, {Recio-Blanco}, {Richards}, {Riello},
  {Rimoldini}, {Robin}, {Roegiers}, {Rybizki}, {Sarro}, {Siopis}, {Smith},
  {Sozzetti}, {Ulla}, {Utrilla}, {van Leeuwen}, {van Reeven}, {Abbas}, {Abreu
  Aramburu}, {Accart}, {Aerts}, {Aguado}, {Ajaj}, {Altavilla}, {{\'A}lvarez},
  {{\'A}lvarez Cid-Fuentes}, {Alves}, {Anderson}, {Anglada Varela}, {Antoja},
  {Audard}, {Baines}, {Baker}, {Balaguer-N{\'u}{\~n}ez}, {Balbinot}, {Balog},
  {Barache}, {Barbato}, {Barros}, {Barstow}, {Bartolom{\'e}}, {Bassilana},
  {Bauchet}, {Baudesson-Stella}, {Becciani}, {Bellazzini}, {Bernet}, {Bertone},
  {Bianchi}, {Blanco-Cuaresma}, {Boch}, {Bombrun}, {Bossini}, {Bouquillon},
  {Bragaglia}, {Bramante}, {Breedt}, {Bressan}, {Brouillet}, {Bucciarelli},
  {Burlacu}, {Busonero}, {Butkevich}, {Buzzi}, {Caffau}, {Cancelliere},
  {C{\'a}novas}, {Cantat-Gaudin}, {Carballo}, {Carlucci}, {Carnerero},
  {Carrasco}, {Casamiquela}, {Castellani}, {Castro-Ginard}, {Castro Sampol},
  {Chaoul}, {Charlot}, {Chemin}, {Chiavassa}, {Cioni}, {Comoretto}, {Cooper},
  {Cornez}, {Cowell}, {Crifo}, {Crosta}, {Crowley}, {Dafonte}, {Dapergolas},
  {David}, {David}, {de Laverny}, {De Luise}, {De March}, {De Ridder}, {de
  Souza}, {de Teodoro}, {de Torres}, {del Peloso}, {del Pozo}, {Delbo},
  {Delgado}, {Delgado}, {Delisle}, {Di Matteo}, {Diakite}, {Diener},
  {Distefano}, {Dolding}, {Eappachen}, {Edvardsson}, {Enke}, {Esquej}, {Fabre},
  {Fabrizio}, {Faigler}, {Fedorets}, {Fernique}, {Fienga}, {Figueras},
  {Fouron}, {Fragkoudi}, {Fraile}, {Franke}, {Gai}, {Garabato},
  {Garcia-Gutierrez}, {Garc{\'\i}a-Torres}, {Garofalo}, {Gavras}, {Gerlach},
  {Geyer}, {Giacobbe}, {Gilmore}, {Girona}, {Giuffrida}, {Gomel}, {Gomez},
  {Gonzalez-Santamaria}, {Gonz{\'a}lez-Vidal}, {Granvik},
  {Guti{\'e}rrez-S{\'a}nchez}, {Guy}, {Hauser}, {Haywood}, {Helmi}, {Hidalgo},
  {Hilger}, {H{\l}adczuk}, {Hobbs}, {Holland}, {Huckle}, {Jasniewicz},
  {Jonker}, {Juaristi Campillo}, {Julbe}, {Karbevska}, {Kervella}, {Khanna},
  {Kochoska}, {Kontizas}, {Kordopatis}, {Korn}, {Kostrzewa-Rutkowska},
  {Kruszy{\'n}ska}, {Lambert}, {Lanza}, {Lasne}, {Le Campion}, {Le Fustec},
  {Lebreton}, {Lebzelter}, {Leccia}, {Leclerc}, {Lecoeur-Taibi}, {Liao},
  {Licata}, {Lindstr{\o}m}, {Lister}, {Livanou}, {Lobel}, {Madrero Pardo},
  {Managau}, {Mann}, {Marchant}, {Marconi}, {Marcos Santos}, {Marinoni},
  {Marocco}, {Marshall}, {Martin Polo}, {Mart{\'\i}n-Fleitas}, {Masip},
  {Massari}, {Mastrobuono-Battisti}, {Mazeh}, {McMillan}, {Messina},
  {Michalik}, {Millar}, {Mints}, {Molina}, {Molinaro}, {Moln{\'a}r},
  {Montegriffo}, {Mor}, {Morbidelli}, {Morel}, {Morris}, {Mulone}, {Munoz},
  {Muraveva}, {Murphy}, {Musella}, {Noval}, {Ord{\'e}novic}, {Orr{\`u}},
  {Osinde}, {Pagani}, {Pagano}, {Palaversa}, {Palicio}, {Panahi}, {Pawlak},
  {Pe{\~n}alosa Esteller}, {Penttil{\"a}}, {Piersimoni}, {Pineau}, {Plachy},
  {Plum}, {Poggio}, {Poretti}, {Poujoulet}, {Pr{\v{s}}a}, {Pulone}, {Racero},
  {Ragaini}, {Rainer}, {Raiteri}, {Rambaux}, {Ramos}, {Ramos-Lerate}, {Re
  Fiorentin}, {Regibo}, {Reyl{\'e}}, {Ripepi}, {Riva}, {Rixon}, {Robichon},
  {Robin}, {Roelens}, {Rohrbasser}, {Romero-G{\'o}mez}, {Rowell}, {Royer},
  {Rybicki}, {Sadowski}, {Sagrist{\`a} Sell{\'e}s}, {Sahlmann}, {Salgado},
  {Salguero}, {Samaras}, {Sanchez Gimenez}, {Sanna}, {Santove{\~n}a},
  {Sarasso}, {Schultheis}, {Sciacca}, {Segol}, {Segovia}, {S{\'e}gransan},
  {Semeux}, {Shahaf}, {Siddiqui}, {Siebert}, {Siltala}, {Slezak}, {Smart},
  {Solano}, {Solitro}, {Souami}, {Souchay}, {Spagna}, {Spoto}, {Steele},
  {Steidelm{\"u}ller}, {Stephenson}, {S{\"u}veges}, {Szabados}, {Szegedi-Elek},
  {Taris}, {Tauran}, {Taylor}, {Teixeira}, {Thuillot}, {Tonello}, {Torra},
  {Torra}, {Turon}, {Unger}, {Vaillant}, {van Dillen}, {Vanel}, {Vecchiato},
  {Viala}, {Vicente}, {Voutsinas}, {Weiler}, {Wevers}, {Wyrzykowski}, {Yoldas},
  {Yvard}, {Zhao}, {Zorec}, {Zucker}, {Zurbach}, \& {Zwitter}}]{Gaia2021}
---. 2021, \aap, 649, A1, \dodoi{10.1051/0004-6361/202039657}

\bibitem[{{Gaia Collaboration} {et~al.}(2022){Gaia Collaboration}, {Vallenari},
  {Brown}, {Prusti}, {de Bruijne}, {Arenou}, {Babusiaux}, {Biermann},
  {Creevey}, {Ducourant}, {Evans}, {Eyer}, {Guerra}, {Hutton}, {Jordi},
  {Klioner}, {Lammers}, {Lindegren}, {Luri}, {Mignard}, {Panem}, {Pourbaix},
  {Randich}, {Sartoretti}, {Soubiran}, {Tanga}, {Walton}, {Bailer-Jones},
  {Bastian}, {Drimmel}, {Jansen}, {Katz}, {Lattanzi}, {van Leeuwen}, {Bakker},
  {Cacciari}, {Casta{\~n}eda}, {De Angeli}, {Fabricius}, {Fouesneau},
  {Fr{\'e}mat}, {Galluccio}, {Guerrier}, {Heiter}, {Masana}, {Messineo},
  {Mowlavi}, {Nicolas}, {Nienartowicz}, {Pailler}, {Panuzzo}, {Riclet}, {Roux},
  {Seabroke}, {Sordo{\o}rcit}, {Th{\'e}venin}, {Gracia-Abril}, {Portell},
  {Teyssier}, {Altmann}, {Andrae}, {Audard}, {Bellas-Velidis}, {Benson},
  {Berthier}, {Blomme}, {Burgess}, {Busonero}, {Busso}, {C{\'a}novas}, {Carry},
  {Cellino}, {Cheek}, {Clementini}, {Damerdji}, {Davidson}, {de Teodoro},
  {Nu{\~n}ez Campos}, {Delchambre}, {Dell'Oro}, {Esquej},
  {Fern{\'a}ndez-Hern{\'a}ndez}, {Fraile}, {Garabato}, {Garc{\'\i}a-Lario},
  {Gosset}, {Haigron}, {Halbwachs}, {Hambly}, {Harrison}, {Hern{\'a}ndez},
  {Hestroffer}, {Hodgkin}, {Holl}, {Jan{\ss}en}, {Jevardat de Fombelle},
  {Jordan}, {Krone-Martins}, {Lanzafame}, {L{\"o}ffler}, {Marchal}, {Marrese},
  {Moitinho}, {Muinonen}, {Osborne}, {Pancino}, {Pauwels}, {Recio-Blanco},
  {Reyl{\'e}}, {Riello}, {Rimoldini}, {Roegiers}, {Rybizki}, {Sarro}, {Siopis},
  {Smith}, {Sozzetti}, {Utrilla}, {van Leeuwen}, {Abbas}, {{\'A}brah{\'a}m},
  {Abreu Aramburu}, {Aerts}, {Aguado}, {Ajaj}, {Aldea-Montero}, {Altavilla},
  {{\'A}lvarez}, {Alves}, {Anders}, {Anderson}, {Anglada Varela}, {Antoja},
  {Baines}, {Baker}, {Balaguer-N{\'u}{\~n}ez}, {Balbinot}, {Balog}, {Barache},
  {Barbato}, {Barros}, {Barstow}, {Bartolom{\'e}}, {Bassilana}, {Bauchet},
  {Becciani}, {Bellazzini}, {Berihuete}, {Bernet}, {Bertone}, {Bianchi},
  {Binnenfeld}, {Blanco-Cuaresma}, {Blazere}, {Boch}, {Bombrun}, {Bossini},
  {Bouquillon}, {Bragaglia}, {Bramante}, {Breedt}, {Bressan}, {Brouillet},
  {Brugaletta}, {Bucciarelli}, {Burlacu}, {Butkevich}, {Buzzi}, {Caffau},
  {Cancelliere}, {Cantat-Gaudin}, {Carballo}, {Carlucci}, {Carnerero},
  {Carrasco}, {Casamiquela}, {Castellani}, {Castro-Ginard}, {Chaoul},
  {Charlot}, {Chemin}, {Chiaramida}, {Chiavassa}, {Chornay}, {Comoretto},
  {Contursi}, {Cooper}, {Cornez}, {Cowell}, {Crifo}, {Cropper}, {Crosta},
  {Crowley}, {Dafonte}, {Dapergolas}, {David}, {David}, {de Laverny}, {De
  Luise}, {De March}, {De Ridder}, {de Souza}, {de Torres}, {del Peloso}, {del
  Pozo}, {Delbo}, {Delgado}, {Delisle}, {Demouchy}, {Dharmawardena}, {Di
  Matteo}, {Diakite}, {Diener}, {Distefano}, {Dolding}, {Edvardsson}, {Enke},
  {Fabre}, {Fabrizio}, {Faigler}, {Fedorets}, {Fernique}, {Fienga}, {Figueras},
  {Fournier}, {Fouron}, {Fragkoudi}, {Gai}, {Garcia-Gutierrez},
  {Garcia-Reinaldos}, {Garc{\'\i}a-Torres}, {Garofalo}, {Gavel}, {Gavras},
  {Gerlach}, {Geyer}, {Giacobbe}, {Gilmore}, {Girona}, {Giuffrida}, {Gomel},
  {Gomez}, {Gonz{\'a}lez-N{\'u}{\~n}ez}, {Gonz{\'a}lez-Santamar{\'\i}a},
  {Gonz{\'a}lez-Vidal}, {Granvik}, {Guillout}, {Guiraud},
  {Guti{\'e}rrez-S{\'a}nchez}, {Guy}, {Hatzidimitriou}, {Hauser}, {Haywood},
  {Helmer}, {Helmi}, {Sarmiento}, {Hidalgo}, {Hilger}, {H{\l}adczuk}, {Hobbs},
  {Holland}, {Huckle}, {Jardine}, {Jasniewicz}, {Jean-Antoine Piccolo},
  {Jim{\'e}nez-Arranz}, {Jorissen}, {Juaristi Campillo}, {Julbe}, {Karbevska},
  {Kervella}, {Khanna}, {Kontizas}, {Kordopatis}, {Korn}, {K{\'o}sp{\'a}l},
  {Kostrzewa-Rutkowska}, {Kruszy{\'n}ska}, {Kun}, {Laizeau}, {Lambert},
  {Lanza}, {Lasne}, {Le Campion}, {Lebreton}, {Lebzelter}, {Leccia}, {Leclerc},
  {Lecoeur-Taibi}, {Liao}, {Licata}, {Lindstr{\o}m}, {Lister}, {Livanou},
  {Lobel}, {Lorca}, {Loup}, {Madrero Pardo}, {Magdaleno Romeo}, {Managau},
  {Mann}, {Manteiga}, {Marchant}, {Marconi}, {Marcos}, {Marcos Santos},
  {Mar{\'\i}n Pina}, {Marinoni}, {Marocco}, {Marshall}, {Polo},
  {Mart{\'\i}n-Fleitas}, {Marton}, {Mary}, {Masip}, {Massari},
  {Mastrobuono-Battisti}, {Mazeh}, {McMillan}, {Messina}, {Michalik}, {Millar},
  {Mints}, {Molina}, {Molinaro}, {Moln{\'a}r}, {Monari}, {Mongui{\'o}},
  {Montegriffo}, {Montero}, {Mor}, {Mora}, {Morbidelli}, {Morel}, {Morris},
  {Muraveva}, {Murphy}, {Musella}, {Nagy}, {Noval}, {Oca{\~n}a}, {Ogden},
  {Ordenovic}, {Osinde}, {Pagani}, {Pagano}, {Palaversa}, {Palicio},
  {Pallas-Quintela}, {Panahi}, {Payne-Wardenaar}, {Pe{\~n}alosa Esteller},
  {Penttil{\"a}}, {Pichon}, {Piersimoni}, {Pineau}, {Plachy}, {Plum}, {Poggio},
  {Pr{\v{s}}a}, {Pulone}, {Racero}, {Ragaini}, {Rainer}, {Raiteri}, {Rambaux},
  {Ramos}, {Ramos-Lerate}, {Re Fiorentin}, {Regibo}, {Richards}, {Rios Diaz},
  {Ripepi}, {Riva}, {Rix}, {Rixon}, {Robichon}, {Robin}, {Robin}, {Roelens},
  {Rogues}, {Rohrbasser}, {Romero-G{\'o}mez}, {Rowell}, {Royer}, {Ruz Mieres},
  {Rybicki}, {Sadowski}, {S{\'a}ez N{\'u}{\~n}ez}, {Sagrist{\`a} Sell{\'e}s},
  {Sahlmann}, {Salguero}, {Samaras}, {Sanchez Gimenez}, {Sanna},
  {Santove{\~n}a}, {Sarasso}, {Schultheis}, {Sciacca}, {Segol}, {Segovia},
  {S{\'e}gransan}, {Semeux}, {Shahaf}, {Siddiqui}, {Siebert}, {Siltala},
  {Silvelo}, {Slezak}, {Slezak}, {Smart}, {Snaith}, {Solano}, {Solitro},
  {Souami}, {Souchay}, {Spagna}, {Spina}, {Spoto}, {Steele},
  {Steidelm{\"u}ller}, {Stephenson}, {S{\"u}veges}, {Surdej}, {Szabados},
  {Szegedi-Elek}, {Taris}, {Taylo}, {Teixeira}, {Tolomei}, {Tonello}, {Torra},
  {Torra}, {Torralba Elipe}, {Trabucchi}, {Tsounis}, {Turon}, {Ulla}, {Unger},
  {Vaillant}, {van Dillen}, {van Reeven}, {Vanel}, {Vecchiato}, {Viala},
  {Vicente}, {Voutsinas}, {Weiler}, {Wevers}, {Wyrzykowski}, {Yoldas}, {Yvard},
  {Zhao}, {Zorec}, {Zucker}, \& {Zwitter}}]{Gaia2022}
{Gaia Collaboration}, {Vallenari}, A., {Brown}, A.~G.~A., {et~al.} 2022, arXiv
  e-prints, arXiv:2208.00211, \dodoi{10.48550/arXiv.2208.00211}

\bibitem[{{Gully-Santiago} {et~al.}(2017){Gully-Santiago}, {Herczeg},
  {Czekala}, {Somers}, {Grankin}, {Covey}, {Donati}, {Alencar}, {Hussain},
  {Shappee}, {Mace}, {Lee}, {Holoien}, {Jose}, \& {Liu}}]{Gully-Santiago2017}
{Gully-Santiago}, M.~A., {Herczeg}, G.~J., {Czekala}, I., {et~al.} 2017, \apj,
  836, 200, \dodoi{10.3847/1538-4357/836/2/200}

\bibitem[{Harris {et~al.}(2020)Harris, Millman, van~der Walt, Gommers,
  Virtanen, Cournapeau, Wieser, Taylor, Berg, Smith, Kern, Picus, Hoyer, van
  Kerkwijk, Brett, Haldane, Fernández~del Río, Wiebe, Peterson,
  Gérard-Marchant, Sheppard, Reddy, Weckesser, Abbasi, Gohlke, \&
  Oliphant}]{Harris2020}
Harris, C.~R., Millman, K.~J., van~der Walt, S.~J., {et~al.} 2020, Nature, 585,
  357–362, \dodoi{10.1038/s41586-020-2649-2}

\bibitem[{{Hayashi}(1961)}]{Hayashi1961}
{Hayashi}, C. 1961, \pasj, 13, 450

\bibitem[{{Hunter}(2007)}]{Hunter2007}
{Hunter}, J.~D. 2007, Computing in Science and Engineering, 9, 90,
  \dodoi{10.1109/MCSE.2007.55}

\bibitem[{{Husser} {et~al.}(2013){Husser}, {Wende-von Berg}, {Dreizler},
  {Homeier}, {Reiners}, {Barman}, \& {Hauschildt}}]{Husser2013}
{Husser}, T.~O., {Wende-von Berg}, S., {Dreizler}, S., {et~al.} 2013, \aap,
  553, A6, \dodoi{10.1051/0004-6361/201219058}

\bibitem[{{Kerr} {et~al.}(2021){Kerr}, {Rizzuto}, {Kraus}, \&
  {Offner}}]{Kerr2021}
{Kerr}, R. M.~P., {Rizzuto}, A.~C., {Kraus}, A.~L., \& {Offner}, S. S.~R. 2021,
  \apj, 917, 23, \dodoi{10.3847/1538-4357/ac0251}

\bibitem[{{Kraus} \& {Hillenbrand}(2008)}]{Kraus2008}
{Kraus}, A.~L., \& {Hillenbrand}, L.~A. 2008, \apjl, 686, L111,
  \dodoi{10.1086/593012}

\bibitem[{{Krolikowski} {et~al.}(2021){Krolikowski}, {Kraus}, \&
  {Rizzuto}}]{Krolikowski2021}
{Krolikowski}, D.~M., {Kraus}, A.~L., \& {Rizzuto}, A.~C. 2021, \aj, 162, 110,
  \dodoi{10.3847/1538-3881/ac0632}

\bibitem[{{Krumholz} {et~al.}(2019){Krumholz}, {McKee}, \&
  {Bland-Hawthorn}}]{Krumholz2019}
{Krumholz}, M.~R., {McKee}, C.~F., \& {Bland-Hawthorn}, J. 2019, \araa, 57,
  227, \dodoi{10.1146/annurev-astro-091918-104430}

\bibitem[{{Lada} \& {Lada}(2003)}]{Lada2003}
{Lada}, C.~J., \& {Lada}, E.~A. 2003, \araa, 41, 57,
  \dodoi{10.1146/annurev.astro.41.011802.094844}

\bibitem[{{Mamajek}(2009)}]{Mamajek2009}
{Mamajek}, E.~E. 2009, in American Institute of Physics Conference Series, Vol.
  1158, Exoplanets and Disks: Their Formation and Diversity, ed. T.~{Usuda},
  M.~{Tamura}, \& M.~{Ishii}, 3--10, \dodoi{10.1063/1.3215910}

\bibitem[{{Morris}(2020)}]{Morris2020}
{Morris}, B.~M. 2020, \apj, 893, 67, \dodoi{10.3847/1538-4357/ab79a0}

\bibitem[{{Paxton} {et~al.}(2011){Paxton}, {Bildsten}, {Dotter}, {Herwig},
  {Lesaffre}, \& {Timmes}}]{Paxton2011}
{Paxton}, B., {Bildsten}, L., {Dotter}, A., {et~al.} 2011, \apjs, 192, 3,
  \dodoi{10.1088/0067-0049/192/1/3}

\bibitem[{{Paxton} {et~al.}(2013){Paxton}, {Cantiello}, {Arras}, {Bildsten},
  {Brown}, {Dotter}, {Mankovich}, {Montgomery}, {Stello}, {Timmes}, \&
  {Townsend}}]{Paxton2013}
{Paxton}, B., {Cantiello}, M., {Arras}, P., {et~al.} 2013, \apjs, 208, 4,
  \dodoi{10.1088/0067-0049/208/1/4}

\bibitem[{{Paxton} {et~al.}(2015){Paxton}, {Marchant}, {Schwab}, {Bauer},
  {Bildsten}, {Cantiello}, {Dessart}, {Farmer}, {Hu}, {Langer}, {Townsend},
  {Townsley}, \& {Timmes}}]{Paxton2015}
{Paxton}, B., {Marchant}, P., {Schwab}, J., {et~al.} 2015, \apjs, 220, 15,
  \dodoi{10.1088/0067-0049/220/1/15}

\bibitem[{{Pecaut} \& {Mamajek}(2016)}]{Pecaut2016}
{Pecaut}, M.~J., \& {Mamajek}, E.~E. 2016, \mnras, 461, 794,
  \dodoi{10.1093/mnras/stw1300}

\bibitem[{{Pecaut} {et~al.}(2012){Pecaut}, {Mamajek}, \& {Bubar}}]{Pecaut2012}
{Pecaut}, M.~J., {Mamajek}, E.~E., \& {Bubar}, E.~J. 2012, \apj, 746, 154,
  \dodoi{10.1088/0004-637X/746/2/154}

\bibitem[{{P{\'e}rez Paolino} {et~al.}(2024){P{\'e}rez Paolino}, {Bary},
  {Hillenbrand}, \& {Markham}}]{PerezPaolino2024}
{P{\'e}rez Paolino}, F., {Bary}, J.~S., {Hillenbrand}, L.~A., \& {Markham}, M.
  2024, arXiv e-prints, arXiv:2403.20255, \dodoi{10.48550/arXiv.2403.20255}

\bibitem[{{P{\'e}rez Paolino} {et~al.}(2023){P{\'e}rez Paolino}, {Bary},
  {Petersen}, {Ward-Duong}, {Tofflemire}, {Follette}, \&
  {Mach}}]{PerezPaolino2023}
{P{\'e}rez Paolino}, F., {Bary}, J.~S., {Petersen}, M.~S., {et~al.} 2023, arXiv
  e-prints, arXiv:2303.01574, \dodoi{10.48550/arXiv.2303.01574}

\bibitem[{{Preibisch} {et~al.}(2002){Preibisch}, {Brown}, {Bridges},
  {Guenther}, \& {Zinnecker}}]{Preibisch2002}
{Preibisch}, T., {Brown}, A. G.~A., {Bridges}, T., {Guenther}, E., \&
  {Zinnecker}, H. 2002, \aj, 124, 404, \dodoi{10.1086/341174}

\bibitem[{{Preibisch} {et~al.}(2001){Preibisch}, {Guenther}, \&
  {Zinnecker}}]{Preibisch2001}
{Preibisch}, T., {Guenther}, E., \& {Zinnecker}, H. 2001, \aj, 121, 1040,
  \dodoi{10.1086/318774}

\bibitem[{{Preibisch} \& {Zinnecker}(1999)}]{Preibisch1999}
{Preibisch}, T., \& {Zinnecker}, H. 1999, \aj, 117, 2381,
  \dodoi{10.1086/300842}

\bibitem[{{Raghavan} {et~al.}(2010){Raghavan}, {McAlister}, {Henry}, {Latham},
  {Marcy}, {Mason}, {Gies}, {White}, \& {ten Brummelaar}}]{Raghavan2010}
{Raghavan}, D., {McAlister}, H.~A., {Henry}, T.~J., {et~al.} 2010, APJS, 190,
  1, \dodoi{10.1088/0067-0049/190/1/1}

\bibitem[{{Rizzuto} {et~al.}(2015){Rizzuto}, {Ireland}, \&
  {Kraus}}]{Rizzuto2015}
{Rizzuto}, A.~C., {Ireland}, M.~J., \& {Kraus}, A.~L. 2015, \mnras, 448, 2737,
  \dodoi{10.1093/mnras/stv207}

\bibitem[{{Salpeter}(1955)}]{Salpeter1955}
{Salpeter}, E.~E. 1955, \apj, 121, 161, \dodoi{10.1086/145971}

\bibitem[{{Somers} {et~al.}(2020){Somers}, {Cao}, \&
  {Pinsonneault}}]{Somers2020}
{Somers}, G., {Cao}, L., \& {Pinsonneault}, M.~H. 2020, \apj, 891, 29,
  \dodoi{10.3847/1538-4357/ab722e}

\bibitem[{{Somers} \& {Pinsonneault}(2015)}]{Somers2015}
{Somers}, G., \& {Pinsonneault}, M.~H. 2015, \apj, 807, 174,
  \dodoi{10.1088/0004-637X/807/2/174}

\bibitem[{{Sullivan} \& {Kraus}(2021)}]{Sullivan2021}
{Sullivan}, K., \& {Kraus}, A.~L. 2021, \apj, 912, 137,
  \dodoi{10.3847/1538-4357/abf044}

\bibitem[{{Tofflemire} {et~al.}(2023){Tofflemire}, {Kraus}, {Mann}, {Newton},
  {Gully-Santiago}, {Vanderburg}, {Waalkes}, {Berta-Thompson}, {Collins},
  {Collins}, {Nielsen}, {Bouchy}, {Ziegler}, {Brice{\~n}o}, \&
  {Law}}]{Tofflemire2023}
{Tofflemire}, B.~M., {Kraus}, A.~L., {Mann}, A.~W., {et~al.} 2023, \aj, 165,
  46, \dodoi{10.3847/1538-3881/aca60f}

\bibitem[{{Tokovinin} \& {Brice{\~n}o}(2020)}]{Tokovinin2020}
{Tokovinin}, A., \& {Brice{\~n}o}, C. 2020, \aj, 159, 15,
  \dodoi{10.3847/1538-3881/ab5525}

\bibitem[{Virtanen {et~al.}(2020)Virtanen, Gommers, Oliphant, Haberland, Reddy,
  Cournapeau, Burovski, Peterson, Weckesser, Bright, {van der Walt}, Brett,
  Wilson, Millman, Mayorov, Nelson, Jones, Kern, Larson, Carey, Polat, Feng,
  Moore, {VanderPlas}, Laxalde, Perktold, Cimrman, Henriksen, Quintero, Harris,
  Archibald, Ribeiro, Pedregosa, {van Mulbregt}, \& {SciPy 1.0
  Contributors}}]{Virtanen2020}
Virtanen, P., Gommers, R., Oliphant, T.~E., {et~al.} 2020, Nature Methods, 17,
  261, \dodoi{10.1038/s41592-019-0686-2}

\bibitem[{{{\v{Z}}erjal} {et~al.}(2023){{\v{Z}}erjal}, {Ireland}, {Crundall},
  {Krumholz}, \& {Rains}}]{Zerjal2023}
{{\v{Z}}erjal}, M., {Ireland}, M.~J., {Crundall}, T.~D., {Krumholz}, M.~R., \&
  {Rains}, A.~D. 2023, \mnras, 519, 3992, \dodoi{10.1093/mnras/stac3693}

\bibitem[{{Ward-Duong} {et~al.}(2015){Ward-Duong}, {Patience}, {De Rosa},
  {Bulger}, {Rajan}, {Goodwin}, {Parker}, {McCarthy}, \&
  {Kulesa}}]{Ward-Duong2015}
{Ward-Duong}, K., {Patience}, J., {De Rosa}, R.~J., {et~al.} 2015, \mnras, 449,
  2618, \dodoi{10.1093/mnras/stv384}

\bibitem[{{Winters} {et~al.}(2019){Winters}, {Henry}, {Jao}, {Subasavage},
  {Chatelain}, {Slatten}, {Riedel}, {Silverstein}, \& {Payne}}]{Winters2019}
{Winters}, J.~G., {Henry}, T.~J., {Jao}, W.-C., {et~al.} 2019, \aj, 157, 216,
  \dodoi{10.3847/1538-3881/ab05dc}

\bibitem[{{Wood} {et~al.}(2021){Wood}, {Mann}, \& {Kraus}}]{Wood2021}
{Wood}, M.~L., {Mann}, A.~W., \& {Kraus}, A.~L. 2021, \aj, 162, 128,
  \dodoi{10.3847/1538-3881/ac0ae9}

\bibitem[{{Zucker} {et~al.}(2022){Zucker}, {Goodman}, {Alves}, {Bialy},
  {Foley}, {Speagle}, {Gro{\^I}{\texttwosuperior}schedl}, {Finkbeiner},
  {Burkert}, {Khimey}, \& {Swiggum}}]{Zucker2022}
{Zucker}, C., {Goodman}, A.~A., {Alves}, J., {et~al.} 2022, \nat, 601, 334,
  \dodoi{10.1038/s41586-021-04286-5}

\end{thebibliography}

\end{document}